\pdfoutput=1

\documentclass[12pt]{article}

\usepackage{tocloft}
\usepackage{graphicx}
\usepackage{natbib}
\usepackage[protrusion=true,expansion=true]{microtype}
\usepackage{amsmath,amsthm,amsfonts,amssymb,mathrsfs,dsfont,mathtools}
\usepackage{booktabs}
\usepackage{url}
\usepackage{float}
\usepackage{color}
\usepackage{parskip}
\usepackage[explicit]{titlesec}
\usepackage{threeparttable}
\usepackage{setspace}
\usepackage[margin=0.85in]{geometry}
\usepackage[font = onehalfspacing]{caption}
\usepackage[usenames,dvipsnames]{xcolor}
\usepackage[hyperfootnotes=true]{hyperref}
\usepackage[labelfont=bf]{caption}
\usepackage{authblk}
\usepackage{caption}
\usepackage[titletoc,toc,title]{appendix}
\usepackage{enumerate}
\usepackage{soul}
\usepackage{subfig}
\usepackage{subfiles} 
\usepackage[capitalise, nameinlink]{cleveref}
\usepackage{dsfont}
\usepackage{chngpage} 

\newcommand{\E}{\mathbb{E}}

\usepackage{scalerel,stackengine}
\stackMath
\newcommand\reallywidehat[1]{%
	\savestack{\tmpbox}{\stretchto{%
			\scaleto{%
				\scalerel*[\widthof{\ensuremath{#1}}]{\kern-.6pt\bigwedge\kern-.6pt}%
				{\rule[-\textheight/2]{1ex}{\textheight}}
			}{\textheight}%
		}{0.5ex}}%
	\stackon[1pt]{#1}{\tmpbox}%
}

\DeclareMathOperator*{\argmin}{arg\,min}

\DeclarePairedDelimiter{\ceil}{\lceil}{\rceil}
\DeclareRobustCommand{\rchi}{{\mathpalette\irchi\relax}}
\newcommand{\irchi}[2]{\raisebox{\depth}{$#1\chi$}} 

\usepackage{graphicx}
\newsavebox\CBox

\makeatletter
\newcommand*\bigcdot{\mathpalette\bigcdot@{.5}}
\newcommand*\bigcdot@[2]{\mathbin{\vcenter{\hbox{\scalebox{#2}{$\m@th#1\bullet$}}}}}
\makeatother

\setstcolor{red} 

\numberwithin{equation}{section}

\hypersetup{colorlinks = true, citecolor = MidnightBlue, linkcolor = MidnightBlue, urlcolor = MidnightBlue}

\titleformat{\section}{\normalfont\large\bfseries}{\thesection}{1em}{#1}
\titleformat{\subsection}{\normalfont\normalsize\bfseries}{\thesubsection}{1em}{#1}
\titleformat{\subsubsection}{\normalfont\normalsize\itshape}{\thesubsubsection}{1em}{#1}
\titlespacing\section{0pt}{12pt plus 4pt minus 2pt}{6pt plus 2pt minus 2pt}
\titlespacing\subsection{0pt}{12pt plus 4pt minus 2pt}{3pt plus 2pt minus 3pt}
\titlespacing\subsubsection{0pt}{12pt plus 4pt minus 2pt}{0pt plus 2pt minus 3pt}

\def\boxit#1{\vbox{\hrule\hbox{\vrule\kern6pt
			\vbox{\kern6pt#1\kern6pt}\kern6pt\vrule}\hrule}}

\definecolor{orange}{rgb}{1,0.5,0}
\definecolor{MyDarkBlue}{rgb}{0,0.08,0.45}

\newtheorem{proposition}{Proposition}[section]
\newtheorem{remark}{Remark}[section]
\newtheorem{theorem}{Theorem}[section]

\newtheorem{assumption}{Assumption}[section]
\newtheorem{Def}{Definition}[section]

\newcommand\blfootnote[1]{%
	\begingroup
	\renewcommand\thefootnote{}\footnote{#1}%
	\addtocounter{footnote}{-1}%
	\endgroup
}

\begin{document}
	
	\title{\Large \bfseries Equal Risk Pricing of Derivatives with Deep Hedging\blfootnote{A GitHub repository with some examples of codes can be found at \href{https://github.com/alexandrecarbonneau}{github.com/alexandrecarbonneau}.}\thanks{
	Alexandre Carbonneau gratefully acknowledges financial support from FRQNT. Fr\'ed\'eric Godin gratefully acknowledges financial support from NSERC (RGPIN-2017-06837).}
	}
	
	\author[a] {Alexandre Carbonneau\thanks{Corresponding author.\vspace{0.2em} \newline
			{\it Email addresses:} \href{mailto:alexandre.carbonneau@mail.concordia.ca}{alexandre.carbonneau@mail.concordia.ca} (Alexandre Carbonneau), \href{mailto:frederic.godin@concordia.ca}{frederic.godin@concordia.ca} (Fr\'ed\'eric Godin).}}
	\author[b]{Fr\'ed\'eric Godin}
	\affil[a,b]{{\small Concordia University, Department of Mathematics and Statistics, Montr\'eal, Canada}}
	
	\vspace{-10pt}
	\date{ 
		\today}
	
	
	\maketitle \thispagestyle{empty} 

	
	\begin{abstract} 
		\vspace{-5pt}
		This article presents a deep reinforcement learning approach to price and hedge financial derivatives. This approach extends the work of \cite{guo2017equal} who recently introduced the equal risk pricing framework, where the price of a contingent claim is determined by equating the optimally hedged residual risk exposure associated respectively with the long and short positions in the derivative. Modifications to the latter scheme are considered to circumvent theoretical pitfalls associated with the original approach. 
		Derivative prices obtained through this modified approach are shown to be arbitrage-free.
		The current paper also presents a general and tractable implementation for the equal risk pricing framework inspired by the deep hedging algorithm of \cite{buehler2019deep}. 
		An $\epsilon$-completeness measure allowing for the quantification of the residual hedging risk associated with a derivative is also proposed. The latter measure generalizes the one presented in \cite{bertsimas2001hedging} based on the quadratic penalty. 
		Monte Carlo simulations are performed under a large variety of market dynamics to demonstrate the practicability of our approach, to perform benchmarking with respect to traditional methods and to conduct sensitivity analyses. 
		
		
		\noindent \textbf{Keywords:} Reinforcement learning, Deep learning, Option pricing, Hedging, Convex risk measures.
		
	\end{abstract} 
	
	\medskip
	
	\thispagestyle{empty} \vfill \pagebreak
	
	\setcounter{page}{1}
	\pagenumbering{roman}
	
	
	
	\doublespacing
	
	\setcounter{page}{1}
	\pagenumbering{arabic}

	
	\section{Introduction}
	Under the complete market paradigm, for instance as in \cite{black1973pricing} and \cite{merton1973theory}, all contingent claims can be perfectly replicated with some dynamic hedging strategy. In such circumstances, the unique arbitrage-free price of an option must be the initial value of the replicating portfolio. However, in reality, markets are incomplete and perfect replication is typically impossible for non-linear derivatives. Indeed, there are many sources of market incompleteness observed in practice such as discrete-time rebalancing, liquidity constraints, stochastic volatility, jumps, etc. In an incomplete market, it is often impracticable for a hedger to select a trading strategy that entirely removes risk as it would typically entail unreasonable costs. For instance, \cite{eberlein1997range} show that the super-replication price of a European call option under a large variety of underlying asset dynamics is the initial underlying asset price. Thus, in practice, a hedger must accept the presence of residual hedging risk that is intrinsic to the contingent claim being hedged. The determination of option prices and hedging policies therefore depend on subjective assumptions regarding risk preferences of market participants.
	
	An incomplete market derivatives pricing approach that is extensively studied in the literature consists in the selection of a suitable equivalent martingale measure (EMM). As shown in the seminal work of \cite{harrison1981martingales}, if a market is incomplete and arbitrage-free, there exists an infinite set of EMMs each of which can be used to price derivatives through a risk-neutral valuation. Some popular examples of EMMs in the literature include the Esscher transform by \cite{Gerber1994} and the minimal-entropy martingale measure by \cite{frittelli2000minimal}. 
	Option pricing functions induced by the latter risk-neutral measures can then be used to calculate Greek letters associated with the option, which leads to the specification of hedging policies, e.g. delta-hedging. However, in that case, hedging policies are not an input of the pricing procedure, but rather a by-product. Thus, hedging policies obtained from many popular EMMs are typically not optimal, and corresponding option prices are not designed in a way that is  consistent with optimal hedging strategies. Another strand of literature derives martingale measures that are designed to be consistent with optimal hedging approaches such as the minimal martingale measure by \cite{follmer1991hedging}, the variance-optimal martingale measure by \cite{schweizer1995variance} 
	and the Extended Girsanov Principle of \cite{elliott1998discrete}. However, an undesirable feature of the three previous methods is their reliance on quadratic objective functions which penalize hedging gains. Moreover, the minimal and variance-optimal martingale measures are often signed measures under realistic models of the underlying asset price, which can be problematic from a theoretical standpoint. The Extended Girsanov Principle, on the contrary, produces a legitimate probability measure. This measure is consistent with a local hedging optimization, i.e. a procedure minimizing incremental discounted risk-adjusted hedging costs. Nevertheless, the identification of a pricing procedure consistent with a non-quadratic global optimization of hedging errors, i.e. a joint optimization over hedging decisions for all time periods until the maturity of the derivative, would be desirable. 

	In that direction, another approach studied in the literature considers the determination of derivatives prices directly from global optimal hedging strategies without having to specify an EMM. 
	A first example of approach among these schemes is utility indifference pricing in which a trader with a specific utility function prices a contingent claim as the value such that the utility of his portfolio remains unchanged by the inclusion of the contingent claim. For instance, \cite{hodges1989optimal} study hedging and indifference pricing under the negative exponential utility function with transaction costs under the Black-Scholes model (BSM). Closely related is the risk indifference pricing in which a risk measure is used to characterize the risk aversion of the trader instead of a utility function. For example, \cite{xu2006risk} studies the indifference pricing and hedging in an incomplete market using convex risk measures as defined in \cite{follmer2002convex}. One notable feature of utility and risk indifference pricing is that the resulting price depends on the position (long or short) of the hedger in the contingent claim. This highlights the need to identify hedging-based pricing schemes producing a unique price that is invariant to being long or short.
	
	Recently, \cite{guo2017equal} introduced the concept of equal risk pricing.
	In their framework, the option price is set as the value such that the global risk exposure of the long and short positions is equal under optimal hedging strategies. 
	Contrarily to utility and risk indifference pricing, equal risk pricing provides a unique transactional price. The latter paper focuses mainly on theoretical features of the equal risk pricing framework and does not provide a general approach to compute the solution of the hedging problem embedded in the methodology. Thus, equal risk prices are only provided for a limited number of specific cases. 
	
	%
	
	To enhance the tractability of the equal risk approach, the current paper considers the use of convex risk measures to quantify the global risk exposures of the long and short positions under optimal hedging strategies. Hedging under a convex risk measure has been extensively studied in the literature: \cite{alexander2003derivative} minimize the Conditional Value-at-Risk (CVaR, \cite{rockafellar2002conditional}) in the context of static hedging with multiple assets, \cite{xu2006risk} studies the indifference pricing and hedging under a convex risk measure in an incomplete market and \cite{godin2016minimizing} develops a global hedging strategy using CVaR as the cost function in the presence of transaction costs. The use of convex risk measures within the equal risk pricing framework was first proposed by \cite{marzban2020equal} who provide a dynamic programming algorithm to compute hedging strategies.
	Recently, \cite{buehler2019deep} introduced an algorithm called deep hedging to hedge a portfolio of over-the-counter derivatives in the presence of market frictions under a convex risk measure using deep reinforcement learning (deep RL). The general framework of RL is for an agent to learn over many iterations of an environment how to select sequences of actions in order to optimize a cost function. In quantitative finance, RL has been applied successfully in algorithmic trading (\cite{moody2001learning}, \cite{lu2017agent} and \cite{deng2016deep}) and in portfolio optimization (\cite{jiang2017deep} and \cite{almahdi2017adaptive}). Hedging with RL also has received some attention. \cite{kolm2019dynamic} demonstrate that SARSA \citep{rummery1994line} can be used to learn the hedging strategy if the objective function is a mean-variance criteria under the BSM. \cite{halperin2017qlbs} shows that Q-learning \citep{watkins1992q} can be used to learn the option pricing and hedging strategy under the BSM. In the novel deep hedging algorithm of \cite{buehler2019deep}, an agent is trained to learn how to optimize the hedging strategy produced by a neural network through many simulations of a synthetic market. Their deep RL approach to the hedging problem helps to counter the well-known curse of dimensionality that arises when the state space gets too large. As argued by \cite{franccois2014optimal}, when applying traditional dynamic programming algorithms to compute hedging strategies, the curse of dimensionality can prevent the use of a large number of features to model the different components of the financial market. 

	The contribution of the current study is threefold. The first contribution consists in providing a universal and tractable methodology to implement the equal risk pricing framework under very general conditions. The approach based on deep RL as in \cite{buehler2019deep} can price and optimally hedge a very large number of contingent claims (e.g. vanilla options, exotic options, options with multiple underlying assets) with multiple liquid hedging instruments under a wide variety of market dynamics (e.g. regime-switching, stochastic volatility, jumps, etc.). Results presented in this paper, which rely on \cite{buehler2019deep}, demonstrate that our methodological approach 
	to equal risk pricing can approximate arbitrarily well the true equal risk price.
	
	The second contribution of the current study consists in performing several numerical experiments studying the behavior of equal risk prices in various contexts. Such experiments showcase the wide applicability of our proposed framework. The behavior of the equal risk pricing approach is analyzed among others through benchmarking against expected risk-neutral pricing and by conducting sensitivity analyzes determining the impact on option prices of the confidence level associated with the risk measure and of the underlying asset model choice. The conduction of such numerical experiments crucially relies on the deep RL scheme outlined in the current study. Using the latter framework allows presenting numerical examples for equal risk pricing that are more extensive, realistic and varied than in previous studies; such results would most likely have been previously inaccessible when relying on more traditional computation methods (e.g. finite difference dynamic programming).
	Numerical results show, among others, that equal risk prices of out-of-the-money (OTM) options are significantly higher than risk-neutral prices across all dynamics considered. This finding is shown to be shared by different option categories which include vanilla and exotic options. Thus, by using the usual risk-neutral valuation instead of the equal risk pricing framework, a risk averse participant trading OTM options might significantly underprice these contracts. 
	
	The last contribution is the introduction of an asymmetric $\epsilon$-completeness measure based on hedging strategies embedded in the equal risk pricing approach. The purpose of the metric is to quantify the magnitude of unhedgeable risk associated with a position in a contingent claim. 
	The $\epsilon$-completeness measure can therefore be used to quantify the level of market incompleteness inherent to a given market model. 
	Our contribution complements the work of \cite{bertsimas2001hedging}; their proposed measure of market incompleteness is based on the mean-squared-error cost function, while ours has the advantage of allowing to characterize the risk aversion of the hedger with any convex risk measure. Furthermore, the current paper's proposed measure is asymmetric in the sense that the risk for the long and short positions in the derivative are quantified by two different hedging strategies, unlike in \cite{bertsimas2001hedging} where the single variance-optimal hedging strategy is considered.
	
	  
	
	The paper is structured as follows. \cref{section_2} introduces our adaptation of the equal risk pricing framework along with the proposed $\epsilon$-completeness measure. \cref{sec:Methodology} describes the deep RL numerical solution to equal risk pricing. \cref{sec:numerical_results} presents various numerical experiments including, among others, sensitivity and benchmarking analyses. \cref{sec:conclusion} concludes. All proofs are provided in \cref{appendix:proofs}.	
	
	
	\section{Equal risk pricing framework}
	\label{section_2}
	
	This section details the theoretical option pricing setup considered in the current study.
	
	\subsection{Market setup}
	\label{subsec:market_description}
	Let $(\Omega, \mathcal{F}, \mathbb{P})$ be the probability space where $\mathbb{P}$ is the physical measure. The financial market is in discrete time with a finite time horizon of $T$ years and known fixed trading dates $\mathcal{T} := \{0 = t_0 < t_{1} < \ldots < t_{N} = T\}$. Consider $D+1$ liquid and tradable assets on the market with $D$ risky assets and one risk-free asset. Risky assets can include for instance stocks and options. Let $\{S_{n}\}_{n=0}^{N}$ be the non-negative price process of the $D$ risky assets where $S_{n} := [S_{n}^{(1)}, \ldots, S_{n}^{(D)}]$ are the prices at time $t_{n} \in \mathcal{T}$. Also, let $\{B_{n}\}_{n=0}^{N}$ be the price process of the risk-free asset with $B_{n} := \text{exp}(rt_{n})$ where $r \in \mathbb{R}$ is the annualized continuously compounded risk-free rate. For convenience, assume that all assets are not paying any cash flows during the trading dates except possibly at time $T$. 
	Define the market filtration $\mathbb{F} := \{\mathcal{F}_{n} \}^N_{n=0}$ where $\mathcal{F}_{n}:=\sigma(S_u|u = 0,\ldots, n), n = 0,\ldots,N$. Moreover, assume that $\mathcal{F} = \mathcal{F}_{N}$. Throughout this paper, suppose that a European-type contingent claim paying off $\Phi(S_{N}, Z_{N}) \geq 0$ 
	at the maturity date $T$ must be priced, where $\{Z_{n}\}^N_{n=0}$ is an $\mathbb{F}$-adapted process with $Z_{n}$ being a $K$-dimensional random vector of relevant state variables and $\Phi : [0,\infty)^{D} \times \mathbb{R}^{K} \rightarrow \mathbb{R}$. 
	$\{Z_{n}\}^N_{n=0}$ can include drivers of risky asset dynamics or information relevant to price the derivative $\Phi$. For the rest of the paper, all assets and contingent claims prices are assumed to be well-behaved and integrable enough. Specific conditions are out-of-scope. 

	Our option pricing approach requires solving the two distinct problems of dynamic optimal hedging, respectively one for a long and one for a short position in the contingent claim. Let $\delta := \{\delta_{n}\}_{n=0}^{N}$ be a trading strategy used by the hedger to minimize his risk exposure to the derivative, where for $n=1,\ldots,N$, $\delta_{n} := [\delta_{n}^{(0)}, \delta_{n}^{(1)}, \ldots, \delta_{n}^{(D)}]$ is a vector containing the number of shares held in each asset during the period $(t_{n-1},t_{n}]$ in the hedging portfolio. $\delta_{n}^{(0)}$ and $\delta_{n}^{(1:D)}:=[\delta_{n}^{(1)}, \ldots, \delta_{n}^{(D)}]$ are respectively the positions in the risk-free asset and in the $D$ risky assets. 
	Furthermore, the initial portfolio (at time $0$ before the first trade) is strictly invested in the risk-free asset. For the rest of the paper, assume the absence of market impact from transactions, i.e. trading in the risky assets does not affect their prices. Here are some well-known definitions in the mathematical finance literature (see for instance \cite{lamberton2011introduction} for more details).
	

	\begin{Def}{(\textit{Discounted gain process})}
		\label{ref:def_disc_gain_process}
		Let $\{G_{n}^{\delta}\}_{n=0}^{N}$ be the discounted gain process associated with the strategy $\delta$ where $G_{n}^{\delta}$ is the discounted gain at time $t_{n}$ prior to the rebalancing. $G_{0}^{\delta} := 0$ and 
		\begin{align}
		G_{n}^{\delta}:= \sum_{k=1}^{n} \delta_{k}^{(1:D)} \bigcdot (B_{k}^{-1} S_{k} - B_{k-1}^{-1} S_{k-1}), \quad n = 1,2,\ldots,N, \label{eq:ref_disc_gain_process_def}
		\end{align}
		where $\bigcdot$ is the dot product operator\footnote{
			If $X = [X_{1},\ldots,X_{n}]$ and $Y = [Y_{1},\ldots,Y_{n}]$, $X \bigcdot Y := \sum_{i=1}^{n}X_{i}Y_{i}$.
		}. 
	\end{Def}


	\begin{Def}{(\textit{Self-financing})}
		\label{ref:def_self_financing}
		The process $\delta$ is said to be a self-financing trading strategy if it is predictable\footnote{
			$X := \{X_{n}\}_{n=0}^{N}$ with $X_{n} := [X_{n}^{(1)},\ldots,X_{n}^{(D)}]$ is $\mathbb{F}$-predictable if for $j = 1,\ldots,D,$ $X_{0}^{(j)} \in \mathcal{F}_{0}$ and $X_{n+1}^{(j)} \in \mathcal{F}_{n}$ for $n = 0,\ldots,N-1$. 
		}
		and if
		\begin{align}
		\delta_{n+1}^{(1:D)} \bigcdot S_{n} + \delta_{n+1}^{(0)}B_{n} = \delta_{n}^{(1:D)} \bigcdot S_{n} + \delta_{n}^{(0)}B_{n}, \quad n = 0,1,\ldots,N-1. \label{eq:ref_self_financing_def}
		\end{align}
		A self-financing strategy $\delta$ implies the absence of cash infusions into or withdrawals from the portfolio except possibly at time $0$.  
	\end{Def}


	\begin{Def}{(\textit{Hedging portfolio value})}
		Define $\{V_{n}^{\delta}\}_{n=0}^{N}$ as the hedging portfolio value process associated with the strategy $\delta$, where the time-$t_n$ portfolio value is given by $V_{n}^{\delta} := \delta_{n}^{(1:D)} \bigcdot S_{n} + \delta_{n}^{(0)}B_{n}$, $n=0,\ldots,N$.
	\end{Def}
	\begin{remark}
		It can be shown, see for instance \cite{lamberton2011introduction}, that $\delta$ is self-financing if and only if $V_{n}^{\delta} = B_{n}(V_{0} + G_{n}^{\delta})$ for $n = 0,1,\ldots,N.$
	\end{remark}
 

	\begin{Def}{(\textit{Admissible trading strategies})}
		\label{ref:def_admissible_set}
		Let $\Pi$ be the convex set of admissible trading strategies which consists of all sufficiently well-behaved self-financing trading strategies. 
	\end{Def}
	

	\subsection{Convex risk measures}
	\label{subsec:convex_risk_measure}
	In an incomplete market, perfect replication is impossible and the hedger must accept that some risks cannot be fully hedged. As such, an optimal hedging strategy (also referred to as a global hedging strategy) is defined as one that minimizes a criterion based on the closeness between the hedging portfolio value and the payoff of the contingent claim at maturity (the difference between two such quantities is referred to as the hedging error). Many different measures of distance can be used to represent the risk aversion of the hedger. In this paper, convex risk measures as defined in \cite{follmer2002convex} are considered. As discussed in \cite{marzban2020equal} and shown in the current section, the use of 
	a convex risk measure to characterize the risk aversion of the hedger enhances the tractability of the equal risk pricing framework.
	\begin{Def}{(\textit{Convex risk measure})} 
		\label{ref:Def_convex_measure}
		Let $\mathcal{X}$ be a set of random variables representing liabilities and $X_{1}, X_{2} \in \mathcal{X}$. As defined in \cite{follmer2002convex}, $\rho : \mathcal{X} \rightarrow \mathbb{R}$ is a convex risk measure if it satisfies the following properties: 
		\begin{enumerate}
			\item[(i)] \textit{Monotonicity}: $X_{1} \leq X_{2} \Rightarrow \rho(X_{1}) \leq \rho(X_{2})$. A larger liability is riskier.
			\item[(ii)] \textit{Translation invariance}: For $a \in \mathbb{R}$, $\rho(X + a) = \rho(X) + a$. This implies the hedger is indifferent between an empty portfolio and a portfolio with a liability $X$ and a cash amount of $\rho(X)$:
			$$\rho(X - \rho(X)) = \rho(X) - \rho(X) = 0.$$
			\item[(iii)] \textit{Convexity}: For $0 \leq \lambda \leq 1$, $\rho(\lambda X_{1} + (1-\lambda)X_{2}) \leq \lambda \rho(X_{1}) + (1-\lambda) \rho(X_{2}).$ Diversification does not increase risk. 
		\end{enumerate}
	\end{Def}


	\subsection{Optimal hedging problem}
	\label{subsec:hedging_problem}
	For the rest of the paper, let $\rho$ be the convex risk measure used to characterize the risk aversion of the hedger for both the long and short positions in the usual contingent claim. Also, assume without loss of generality (w.l.o.g.) that the position in the hedging portfolio is long for both the long and short positions in the derivative.
	\begin{Def}{(\textit{Long and short sided risk)}}
		\label{HedgingProbDef}
		Define $\epsilon^{(L)}(V_0)$ and $\epsilon^{(S)}(V_0)$ respectively as the measured risk exposure of a long and short position in the derivative under the optimal hedge if the value of the initial hedging portfolio is $V_{0}:$
		\begin{align}
		\epsilon^{(L)}(V_0) &:= \underset{\delta\in \Pi}{\min} \, \rho \left(-\Phi(S_{N},Z_{N}) -B_{N}(V_0 + G_{N}^{\delta})\right), \label{eq:risk_long}
		\\ \epsilon^{(S)}(V_0) &:= \underset{\delta \in \Pi}{\min} \, \rho \left(\Phi(S_{N},Z_{N}) - B_{N}(V_0 + G_{N}^{\delta})\right). \label{eq:risk_short}
		\end{align}
	\end{Def} 
	\begin{remark}
		\label{remark:inf_vs_min}
		An assumption implicit to \cref{HedgingProbDef} is that the minimum in \eqref{eq:risk_long} or \eqref{eq:risk_short} is indeed attained by some trading strategy, i.e. that the infimum is in fact a minimum. The identification of conditions which ensure that this assumption is satisfied are left out-of-scope.
	\end{remark}
	
	We emphasize that the optimal risk exposures of the long and short position as defined in \eqref{eq:risk_long} and \eqref{eq:risk_short} are reached through two distinct hedging strategies. The following proposition is a direct consequence of the translation invariance of $\rho$.
	\begin{proposition}
		\label{prop:risk_exposure}
		\begin{align}
		\epsilon^{(L)}(V_0) &= \epsilon^{(L)}(0) - B_{N}V_{0}, \quad \epsilon^{(S)}(V_0) = \epsilon^{(S)}(0) - B_{N}V_{0}. \nonumber
		\end{align} 
	\end{proposition}

	
	\begin{Def}{(\textit{Optimal hedging})}
		Let $\delta^{(L)}$ and $\delta^{(S)}$ be respectively the optimal hedging strategies for the long and short positions in the derivative:
		\begin{align}
		\delta^{(L)}&:= \underset{\delta \in \Pi}{\argmin} \, \rho \left(-\Phi(S_{N},Z_{N}) -B_{N}(V_0 + G_{N}^{\delta})\right), \label{eq:ref_optimal_long_def}
		\\ \delta^{(S)}&:= \underset{\delta \in \Pi}{\argmin} \, \rho \left(\Phi(S_{N},Z_{N}) - B_{N}(V_{0} + G_{N}^{\delta})\right). \label{eq:ref_optimal_short_def}
		\end{align}  
	\end{Def}

	
	The translation invariance property of $\rho$ implies that the optimal hedging strategies $\delta^{(L)}$ and $\delta^{(S)}$ 
	do not depend on $V_{0}$, as stated in the following proposition.
	\begin{proposition}[\textit{Independence of the optimal hedging strategies from $V_{0}$}]
		\label{prop:ind_V0_hedge}		 
		\begin{align}
		\delta^{(L)} &= \underset{\delta \in \Pi}{\argmin} \, \rho \left(-\Phi(S_{N},Z_{N}) -B_{N}G_{N}^{\delta}\right), \label{eq:ref_optimal_long_ind_V0}
		\\ \delta^{(S)} &= \underset{\delta \in \Pi}{\argmin} \, \rho \left(\Phi(S_{N},Z_{N}) - B_{N}G_{N}^{\delta}\right). \label{eq:ref_optimal_short_ind_V0}
		\end{align}
	\end{proposition}


	\subsection{Option pricing and $\epsilon$-completeness measure}
	\label{subsec:option_pricing_incomplete_measure}	
	The current section outlines the equal risk pricing criterion to determine the price of a derivative. It entails finding a price for which the risk exposure to both the long position and short position hedgers are equal. One important concept in the valuation of contingent claims is the absence of arbitrage. In this paper, the notions of super-replication and sub-replication are used to define arbitrage-free pricing.
	\begin{Def}[\textit{Super-replication and sub-replication strategies}]
		\label{def:super_sub_strat}
		A super-replication strategy for the contingent claim $\Phi$ is defined as a pair $(v, \delta)$ such that $v\in\mathbb{R}$, and $\delta$ is an admissible hedging strategy for which $V^\delta_0 = v$ and $V_{N}^{\delta}=B_{N}(v + G_{N}^{\delta}) \geq \Phi(S_{N},Z_{N})$ $\mathbb{P}$-a.s. Super-replication is a conservative approach to hedging which can be used by a seller of $\Phi$ to remove all residual hedging risk. Let $\bar{v}$ be the greatest lower bound of the set of initial portfolio values for which a super-replication strategy exists:
		\begin{align}
		\bar{v}&:= \inf \left\{v: \exists \delta \in \Pi \text{ such that }  \mathbb{P} \left[ B_{N}(v + G_{N}^{\delta}) \geq \Phi(S_{N},Z_{N}) \right] = 1 \right\}. \label{eq:ref_super_replication_price}
		\end{align}
		$\bar{v}$ is called the super-replication price of $\Phi$ and it represents an upper bound of the set of arbitrage-free prices for $\Phi$. Similarly, a sub-replication strategy is a pair $(v, \delta)$ that completely removes the hedging risk exposure associated with a long position in $\Phi$, i.e. for which $B_{N}(v + G_{N}^{\delta}) \leq \Phi(S_{N},Z_{N})$ $\mathbb{P}$-a.s. The least upper bound of the set of portfolio values such that a sub-replication strategy exists is called the sub-replication price of $\Phi$ and is a lower bound of the set of arbitrage-free prices for $\Phi$:
		\begin{align}
		\underline{v}&:= \sup \left\{v: \exists \delta \in \Pi \text{ such that } \mathbb{P} \left[ B_{N}(v + G_{N}^{\delta}) \leq \Phi(S_{N},Z_{N}) \right] = 1 \right\}. \label{eq:ref_sub_replication_price}
		\end{align}
		
	\end{Def}

	\begin{Def}[{Arbitrage-free pricing}]
	\label{def:arb_free_pricing}
	For the rest of the paper, the price of a contingent claim $\Phi$ is said to be arbitrage-free if it falls within the interval $[\underline{v}, \bar{v}]$ as defined in \eqref{eq:ref_super_replication_price} and \eqref{eq:ref_sub_replication_price}. 
	\end{Def}


	The price of a contingent claim under the equal risk pricing framework can now be defined.
	
	\begin{Def}[\textit{Equal risk price for European-type claims}]
		\label{def:equal_risk_price}
		The equal risk price $C_{0}^{\star}$ of the contingent claim $\Phi$ is defined as the initial portfolio value such that the optimally hedged measured risk exposure of both the long and short positions in the derivative are equal, i.e. $C_{0}^{\star}:=C_{0}$ such that:
		\begin{align}
		\epsilon^{(L)}(-C_{0}) = \epsilon^{(S)}(C_{0}). \label{eq:ref_what_ever}
		\end{align} 
	\end{Def}
	\begin{remark}
		Contrarily to \cite{guo2017equal}, in the current paper, the optimal hedging strategy minimizes risk under the physical measure instead of under some risk-neutral measure. Two main reasons led to this modification of the original approach found in \cite{guo2017equal}. First, under incomplete markets, the choice of the risk-neutral measure is arbitrary, whereas the physical measure can be more objectively determined using econometrics techniques. Having the price being determined under the physical measure removes the subjectivity associated with the choice of the martingale measure. Secondly, under a risk-neutral measure $\mathbb{Q}$, the price is already characterized by the discounted expected payoff $\mathbb{E}^\mathbb{Q} \left[e^{-rT} \Phi(S_N,Z_N) \right]$, which makes interpretation of the risk-neutral equal risk price questionable. 
	\end{remark}
	
	Before introducing results showing that equal risk option prices are arbitrage-free, a technical assumption on which the proofs rely is outlined.
	\begin{assumption}
		As in \cite{xu2006risk} and \cite{marzban2020equal}, assume that the risk associated to hedging losses is bounded below across all admissible trading strategies, i.e. $\underset{\delta \in \Pi}{\min} \ \rho (-B_{N}G_{N}^{\delta})> - \infty$.
	\end{assumption}
	The next theorem provides a characterization of equal risk prices. It also indicates that equal risk prices of contingent claims with a finite super-replication price are arbitrage-free. The representation of $C_{0}^{\star}$ in \eqref{eq:ref_equalriskprice} is analogous to results found in \cite{marzban2020equal} who considers a similar setup with convex risk measures. Although the arbitrage-free result is also stated in \cite{marzban2020equal}, a formal proof was not given.
	\begin{theorem}[\textit{Absence of arbitrage}]
	\label{theorem:uniqueness_equal_risk_price}
	 Assume that there exist a finite super-replication price for $\Phi$. Then, the equal risk price $C_{0}^{\star}$ from \cref{def:equal_risk_price} exists, is unique, is arbitrage-free and can be expressed as
		\begin{align}
		C_{0}^{\star} &= \frac{\epsilon^{(S)}(0) - \epsilon^{(L)}(0)}{2 B_{N}}.  \label{eq:ref_equalriskprice}
		\end{align}
	\end{theorem}

	\begin{remark}
	The risk measure considered in the work of \cite{guo2017equal} lacks the translation invariance property, which implies that equal risk prices are provided for a very limited number of cases. The proof of \cref{theorem:uniqueness_equal_risk_price} shows that the representation of $C_{0}^{\star}$ in \eqref{eq:ref_equalriskprice} is a direct consequence of the translation invariance property of convex risk measures.
	\end{remark}

	We now propose measures to quantify the residual risk faced by hedgers of the contingent claim. Such measures are analogous to but more general than the one proposed in \cite{bertsimas2001hedging} who study the case of variance-optimal hedging.
	\begin{Def}[\textit{$\epsilon$ market completeness measure}]
		\label{ref:def_epsilon_complete_measure}
		Define $\epsilon^{\star}$ as the level of residual risk faced by the hedger of any of the short or long position in the contingent claim if its price is the equal risk price and optimal hedging strategies are used for both positions:
		\begin{align}
		\epsilon^{\star} := \epsilon^{(L)}(-C_{0}^{\star}) = \epsilon^{(S)}(C_{0}^{\star}). \label{eq:ref_equalrisk_measure} 
		\end{align}	
		$\epsilon^{\star}$ and $\epsilon^{\star}/C_{0}^{\star}$ are referred to as respectively the measured residual risk exposure per derivative contract and per dollar invested.
	\end{Def}
	The following proposition states that $\epsilon^{\star}$ is the average of the measured risk exposure of both long and short optimally hedged positions in the contingent claim $\Phi$ assuming that the initial value of the portfolio is zero. 
	\begin{proposition}
		\label{proposition:interesting_form_risk_exposure}
		\begin{align}
		\epsilon^{\star} = \frac{\epsilon^{(L)}(0) + \epsilon^{(S)}(0)}{2}. \label{eq:ref_epsilon_star_mean}
		\end{align}
	\end{proposition}

	
	\begin{remark}
		\cite{bertsimas2001hedging} proposed instead the following measure of market incompleteness:
		$$\varepsilon^{\star} = \underset{V_{0}, \delta}{\min} \, \mathbb{E}[(\Phi(S_{N},Z_{N})- B_{N}(V_{0} + G_{N}^{\delta}) )^{2}],$$
		where the expectation is taken with respect to the physical measure.
		Our measure $\epsilon^{\star}$ has the advantage of characterizing the risk aversion of the hedger with a convex risk measure, contrarily to \cite{bertsimas2001hedging} who are restricted to the use of a quadratic penalty.
		Using the latter penalty entails that hedging gains are penalized during the optimization of the hedging strategy, which is clearly undesirable. The ability to rely on convex measures in the current scheme for risk quantification allows for an asymmetric treatment of hedging gains and losses which is more consistent of actual objectives of the hedging agents.
	\end{remark}
	
	As argued by \cite{bertsimas2001hedging}, market incompleteness is often described in the literature as a binary concept whereas in practice, it is much more natural to consider different degrees of incompleteness implying different levels of residual hedging risk. The measure $\epsilon^{\star}$ allows determining where is any contingent claim situated within the spectrum of incompleteness and whether it is easily hedgeable or not. As discussed in \cite{bertsimas2001hedging}, a single metric such as $\epsilon^{\star}$ might not be sufficient for a complete depiction of the level of market incompleteness associated with a contingent claim. For instance, it does not depict the entire hedging error distribution, nor does it directly indicate which scenarios are the main drivers of hedging residual risk. Nevertheless, $\epsilon^{\star}$ is still a good indication of the efficiency of the optimal hedging procedure for a given derivative. Moreover, sensitivity analyses over $\epsilon^{\star}$ with respect to various model dynamics can be done to assess the impact of the different sources of market incompleteness. Numerical experiments in \cref{sec:numerical_results} will attempt to provide some insight on drivers of $\epsilon^{\star}$.

	\section{Tractable solution to equal risk pricing}
	\label{sec:Methodology}
	In the current section, a tractable solution is proposed to implement the equal risk pricing framework. The approach uses the recent deep hedging algorithm of \cite{buehler2019deep} to train two distinct neural networks which are used to approximate the optimal hedging strategy respectively for the long and the short position in the derivative.
	
	
	\subsection{Feedforward neural network} 
	\label{subsec:FFNN}
	For convenience, a very similar notation for neural networks as the one introduced by \cite{buehler2019deep} is used (see Section $4$ of their paper). The reader is referred to \cite{goodfellow2016deep} for a general description of neural networks.
	
	\begin{Def}[\textit{Feedforward neural network}]
		\label{def:ref_FFNN}
		Let $X \in \mathbb{R}^{d_{\text{in}} \times 1}$ be a feature vector of dimensions $d_{\text{in}} \in \mathbb{N}$ and $L, d_{1},\ldots, d_{L-1}, d_{\text{out}} \in \mathbb{N}$ with $L \geq 2$. Define a feedforward neural network (FFNN) as the mapping $F_{\theta}:\mathbb{R}^{d_{\text{in}}} \rightarrow \mathbb{R}^{d_{\text{out}}}$ with trainable parameters $\theta$:
		\begin{align}
		F_{\theta}(X) &:= A_{L} \circ F_{L-1} \circ \ldots \circ F_{1}, \label{eq:ref_FFNN}
		\\ F_{l} &:= \sigma \circ A_{l}, \quad l = 1,\ldots, L-1, \nonumber
		\end{align}
		where $\circ$ denotes the function composition operator, and for any $l=1,\ldots, L$, the function $A_{l}$ is defined through $A_{l}(Y):= W^{(l)}Y + b^{(l)}$ with
		\begin{itemize}
		\item $W^{(1)} \in \mathbb{R}^{d_{1} \times d_{\text{in}}}$, $b^{(1)} \in \mathbb{R}^{d_1 \times 1}$ and $Y \in \mathbb{R}^{d_{\text{in}} \times 1}$ if $l =1$,
		\item $W^{(l)} \in \mathbb{R}^{d_{l} \times d_{l-1}}$, $b^{(l)} \in \mathbb{R}^{d_l \times 1}$ and $Y \in \mathbb{R}^{d_{l-1} \times 1}$ if $l = 2,\ldots,L-1$ and $L\geq 3$,
		\item $W^{(L)} \in \mathbb{R}^{d_{\text{out}} \times d_{L-1}}$, $b^{(L)} \in \mathbb{R}^{d_{\text{out}} \times 1}$ and $Y \in \mathbb{R}^{d_{L-1} \times 1}$ if $l = L$.
		\end{itemize}
		 The activation function $\sigma: \mathbb{R} \rightarrow \mathbb{R}$ is applied element-wise to outputs of the pre-activation functions $A_{l}$. 
		Moreover, 
		\begin{align}
		\theta := \{W^{(1)}, \ldots, W^{(L)}, b^{(1)}, \ldots, b^{(L)}\} \label{eq:ref_train_params}
		\end{align} 
		is the set of trainable parameters of the FFNN.
	\end{Def}
	
	
	The following definition of sets of FFNN will be used throughout the rest of the section to define, for instance, the two neural networks used for hedging the long and short position in the derivative, the tractable solution to the equal risk pricing framework and the optimization procedure of neural networks.
	
	
	\begin{Def}[\textit{Sets of FFNN}]
	\label{ref:def_sets_FFNN}
	Let $\mathcal{NN}_{\infty, d_{\text{in}}, d_{\text{out}}}^{\sigma}$ be the set of all FFNN mapping from $\mathbb{R}^{d_{\text{in}}} \rightarrow \mathbb{R}^{d_{\text{out}}}$ as in \cref{def:ref_FFNN} with a fixed activation function $\sigma$ and an arbitrary number of layers and neurons per layer. Since a unique activation function is considered in the numerical section, let $\mathcal{NN}_{\infty, d_{\text{in}}, d_{\text{out}}}:= \mathcal{NN}_{\infty, d_{\text{in}}, d_{\text{out}}}^{\sigma}$. Moreover, for  all $M \in \mathbb{N}$ and $R_M \in \mathbb{N}$ that depends on $M$, let $\Theta_{M, d_{\text{in}}, d_{\text{out}}} \subseteq \mathbb{R}^{R_M}$. 
	Define $\mathcal{NN}_{M, d_{\text{in}}, d_{\text{out}}}$ as the set of neural networks $F_{\theta}$ as in \eqref{eq:ref_FFNN} with $\theta \in \Theta_{M, d_{\text{in}}, d_{\text{out}}}$:
	\begin{align}
	\mathcal{NN}_{M, d_{\text{in}}, d_{\text{out}}} := \{F_{\theta} : \theta \in \Theta_{M, d_{\text{in}}, d_{\text{out}}}\}. \label{eq:ref_sets_of_FFNN} 
	\end{align}
	The sequence of sets $\{\mathcal{NN}_{M, d_{\text{in}}, d_{\text{out}}}\}_{M \in \mathbb{N}}$ is assumed to have the following properties:
	\begin{itemize}
		\item For any $M \in \mathbb{N}$: $\mathcal{NN}_{M, d_{\text{in}}, d_{\text{out}}} \subset \mathcal{NN}_{M+1, d_{\text{in}}, d_{\text{out}}}$ where $\subset$ denotes strict inclusion,
		\item $\bigcup_{M \in \mathbb{N}} \mathcal{NN}_{M, d_{\text{in}}, d_{\text{out}}} = \mathcal{NN}_{\infty, d_{\text{in}}, d_{\text{out}}}$.
	\end{itemize}
	\end{Def}
	
	
	This definition of sets of FFNN introduced by \cite{buehler2019deep} is very convenient as the sets $\{\mathcal{NN}_{M, d_{\text{in}}, d_{\text{out}}}\}_{N \in \mathbb{N}}$ can be used to describe two cases of interest in deep learning. Here are two different possible definitions for $\mathcal{NN}_{M, d_{\text{in}}, d_{\text{out}}}$.
	\begin{itemize}
		\item [(A)] Let $\{L^{(M)}\}_{M \in \mathbb{N}}$, $\{d_{1}^{(M)}\}_{M \in \mathbb{N}}, \{d_{2}^{(M)}\}_{M \in \mathbb{N}}, \ldots$ be non-decreasing integer sequences. 
		Then, $\mathcal{NN}_{M, d_{\text{in}}, d_{\text{out}}}$ is defined as the set of all FFNN mapping from $\mathbb{R}^{d_{\text{in}}} \rightarrow \mathbb{R}^{d_{\text{out}}}$ with a fixed structure of $L^{(M)}$ layers and of $d_{1}^{(M)},\ldots,d_{L^{(M)}-1}^{(M)}, d_{\text{out}}$ neurons per layer. This case is useful for the problem of fitting the trainable parameters $\theta$ with a fixed set of hyperparameters. 
		\item [(B)] Let $\mathcal{NN}_{M, d_{\text{in}}, d_{\text{out}}}$ be the set of all FFNN mapping from $\mathbb{R}^{d_{\text{in}}} \rightarrow \mathbb{R}^{d_{\text{out}}}$ for an arbitrary number of layers and number of neurons per layer with at most $M$ non-zero trainable parameters. This case is useful to describe the complete optimization problem of neural networks which include the selection of hyperparameters, often called hyperparameters tuning. 
	\end{itemize}
	Unless specified otherwise, one can assume w.l.o.g. either definition for $\{\mathcal{NN}_{M, d_{\text{in}}, d_{\text{out}}}\}_{M \in \mathbb{N}}$.
	
	
	\subsection{Equal risk pricing with two neural networks}
	\label{subsec:option_pricing_and_hedging_with_FFNN}
	To formulate how two distinct neural networks can approximate arbitrarily well the optimal hedging of the long and short position in a derivative, the following assumption is applied for the rest of the paper.
	
	
	\begin{assumption}
		\label{assumpt:unique}
		For each position (long and short) in the derivative, there exists a function $f:\mathbb{R}^{\tilde{D}} \rightarrow \mathbb{R}^{D}$ (distinct for the long and short position) such that at each rebalancing date, the optimal hedge is of the form $\delta_{n+1}^{(1:D)} = f(S_{n}, V_{n}, \mathcal{I}_{n}, T - t_{n})$ where $\tilde{D}:=D+2+\text{dim}(\mathcal{I}_n)$ with $\mathcal{I}_n$ being some random vector encompassing relevant necessary information to compute the optimal hedging strategy, which depends on the market setup considered. 
	\end{assumption}
	
	
	Note that \cref{assumpt:unique} typically holds for low-dimension processes $\{ \mathcal{I}_n \}^N_{n=0}$ when some form of Markov dynamics common in the hedging literature is assumed. See, for example, \cite{franccois2014optimal} for the case of regime-switching models. 
	
	In what follows, $\mathcal{L}$ and $\mathcal{S}$ used both as subscripts and superscripts denote respectively the long and short position hedges.
	
	
	\begin{Def}[\textit{Hedging with two neural networks}] 
		\label{def:long_short_FFNN}
		Let $X_{n}:=(S_{n}, V_{n}, \mathcal{I}_{n}, T-t_{n}) \in \mathbb{R}^{\tilde{D}}$ be the feature vector for each trading time $t_{n} \in \{t_{0}, \ldots, t_{N-1}\}$. For some $M_{\mathcal{L}} \in \mathbb{N}$, let $F_{\theta}^{\mathcal{L}} \in \mathcal{NN}_{M_{\mathcal{L}}, \tilde{D}, D}$ be a FFNN.
		Given $X_{n}$ as an input, $F_{\theta}^{\mathcal{L}}$ outputs a $D$-dimensional vector of the number of shares of each of the $D$ risky assets held in the hedging portfolio of the long position during the period $(t_{n}, t_{n+1}]$, i.e. $\delta_{n+1}^{(1:D)} = F_{\theta}^{\mathcal{L}}(X_{n})$. Similarly, for some $M_{\mathcal{S}} \in \mathbb{N}$, $F_{\theta}^{\mathcal{S}} \in \mathcal{NN}_{M_{\mathcal{S}}, \tilde{D}, D}$ is a distinct FFNN which computes the position in the $D$ risky assets to hedge the short position in the option at each time step. These two FFNN are referred to as the long-$\mathcal{NN}$ and short-$\mathcal{NN}$. 
	\end{Def}

	
	\begin{remark}
		In the current paper's approach, the two neural networks are trained separately to minimize different cost functions. As such, $F_{\theta}^{\mathcal{L}}$ and $F_{\theta}^{\mathcal{S}}$ will possibly have a different structure, e.g. different number of layers and number of neurons per layer, and different values of trainable parameters.
	\end{remark}

	 The problem of evaluating the measured risk exposure of the long and short positions under optimal hedging can now be formulated as a classical deep learning optimization problem. Since the input and output of $F_{\theta}^{\mathcal{L}}$ and $F_{\theta}^{\mathcal{S}}$ are always respectively of dimensions $\tilde{D}$ and $D$, let $\Theta_{M_{\mathcal{L}}}:=\Theta_{M_{\mathcal{L}, \tilde{D}, D}}$ and $\Theta_{M_{\mathcal{S}}}:=\Theta_{M_{\mathcal{S}, \tilde{D}, D}}$ be the sets of trainable parameters values as in \eqref{eq:ref_sets_of_FFNN} for respectively the long-$\mathcal{NN}$ and short-$\mathcal{NN}$.
	
	
	\begin{Def}[\textit{Long and short sided risk with two neural networks}]
		\label{eq:def_long_short_risk_NN}
		For $M_{\mathcal{L}}, M_{\mathcal{S}} \in \mathbb{N}$, define $\epsilon^{(M_{\mathcal{L}})}(V_0)$ and $\epsilon^{(M_\mathcal{S})}(V_0)$ as the measured risk exposure of the long and short position in the derivative if $F_{\theta}^{\mathcal{L}}$ and $F_{\theta}^{\mathcal{S}}$ are used to compute the hedging strategies and the initial hedging portfolio value is $V_{0}$: 
		\begin{align}
		\epsilon^{(M_{\mathcal{L}})}(V_0) &:= \underset{\theta \in \Theta_{M_{\mathcal{L}}}}{\min} \, \rho \left(- \Phi(S_{N},Z_{N}) -B_{N}(V_0 + G_{N}^{\delta^{(L, \theta)}})\right), \label{eq:ref_risk_long_FFNN}
		\\ \epsilon^{(M_{\mathcal{S}})}(V_0) &:= \underset{\theta \in \Theta_{M_{\mathcal{S}}}}{\min} \, \rho \left(\Phi(S_{N},Z_{N}) - B_{N}(V_0 + G_{N}^{\delta^{(S, \theta)}})\right), \label{eq:ref_risk_short_FFNN}
		\end{align}
		where $\delta^{(L, \theta)}$ and $\delta^{(S, \theta)}$ in \eqref{eq:ref_risk_long_FFNN} and \eqref{eq:ref_risk_short_FFNN} are to be understood respectively as the trading strategies obtained through $F_{\theta}^{\mathcal{L}}$ and $F_{\theta}^{\mathcal{S}}$.
	\end{Def}
	
	
	\begin{remark}
		Following similar steps as in the proof of \cref{prop:risk_exposure}, it can be shown that
		\begin{align}
		\epsilon^{(M_{\mathcal{L}})}(V_0) = \epsilon^{(M_{\mathcal{L}})}(0) - B_{N}V_{0}, \quad \epsilon^{(M_\mathcal{S})}(V_0) = \epsilon^{(M_\mathcal{S})}(0) - B_{N}V_{0}. \nonumber
		\end{align} 
	\end{remark} 
	
		
	\begin{remark}
		\label{remark:optimal_hedging_convergence}
		Suppose \cref{assumpt:unique} is satisfied. 
		Using the universal function approximation theorem of \cite{hornik1991approximation} which essentially states that a FFNN approximates multivariate functions arbitrarily well, \cite{buehler2019deep} show that for any well-behaved and integrable enough asset prices dynamics and contingent claims (see Proposition 4.3 of their paper\footnote{
			\cite{buehler2019deep} consider a more general market with a filtration generated by a process $\{I_{k}\}$ where $I_{k} \in \mathbb{R}^{d}$ contains any new market information at time $t_{k}$. They use a distinct neural network at each trading date which can be a function of $(I_{0},\ldots, I_{k},\delta_{k})$ at time $t_{k}$. From remarks $5$ and $6$ of \cite{buehler2019deep}, the convergence result \eqref{eq:ref_convergence} holds under \cref{assumpt:unique} by using instead a single FFNN for both the long and short position for all time steps as in \cref{def:long_short_FFNN} of the current paper.
		}):
		\begin{align}
		\lim\limits_{M_\mathcal{S} \rightarrow \infty} \epsilon^{(M_\mathcal{S})}(0) = \epsilon^{(S)}(0), \quad \lim\limits_{M_{\mathcal{L}} \rightarrow \infty} \epsilon^{(M_{\mathcal{L}})}(0) = \epsilon^{(L)}(0). \label{eq:ref_convergence}
		\end{align}
		Thus, this result shows that for both the long and short positions, there exists a large FFNN which can approximate arbitrarily well the optimal hedging strategy. 
	\end{remark}

	
	The equal risk pricing approach as well as the measure of market incompleteness $\epsilon$ can now be restated with the use of the long-$\mathcal{NN}$ and short-$\mathcal{NN}$.
	\begin{Def}[\textit{Equal risk pricing and $\epsilon$-completeness measure with two neural networks}]
		\label{def:equal_risk_with_two_FFNN}
		Define $C_{0}^{(\star, \mathcal{NN})}$ as the equal risk price if $F_{\theta}^{\mathcal{L}}$ and $F_{\theta}^{\mathcal{S}}$ are used to compute the hedging strategies, i.e. $C_{0}^{(\star, \mathcal{NN})}:=C_{0}$ such that:
		$$\epsilon^{(M_{\mathcal{L}})}(-C_{0}) = \epsilon^{(M_\mathcal{S})}\left(C_{0}\right).$$
		Furthermore, let $\epsilon^{(\star, \mathcal{NN})}$ be the measure of market incompleteness if the price of the derivative is $C_{0}^{(\star, \mathcal{NN})}$:
		\begin{align}
		\epsilon^{(\star, \mathcal{NN})} &:= \epsilon^{(M_{\mathcal{L}})}\left(-C_{0}^{(\star, \mathcal{NN})}\right) = \epsilon^{(M_\mathcal{S})}\left(C_{0}^{(\star, \mathcal{NN})}\right). \label{eq:ref_equalrisk_measure_with_FFNN} 
		\end{align}
	\end{Def}

	
	\begin{remark}
		Following similar steps as in the proof of \cref{theorem:uniqueness_equal_risk_price} and \cref{proposition:interesting_form_risk_exposure}, it can be shown that
		\begin{align}
		C_{0}^{(\star, \mathcal{NN})} = \frac{\epsilon^{(M_\mathcal{S})}(0) - \epsilon^{(M_{\mathcal{L}})}(0)}{2 B_{N}}, \quad \epsilon^{(\star, \mathcal{NN})} = \frac{\epsilon^{(M_{\mathcal{L}})}(0) + \epsilon^{(M_\mathcal{S})}(0)}{2}. \label{eq:ref_equal_risk_price_measure_form}
		\end{align} 
	\end{remark}

	
	An important consequence of \cref{remark:optimal_hedging_convergence} is that the current paper's approach based on neural networks can approximate arbitrarily well the true equal risk price and measure of incompleteness. 
	\begin{proposition}[]
		\label{prop_ref_arbitrage_free_with_NN}
		\begin{align}
		\lim\limits_{M_\mathcal{S}, M_{\mathcal{L}} \rightarrow \infty}C_{0}^{(\star, \mathcal{NN})} &= C_{0}^{\star}, \quad \lim\limits_{M_\mathcal{S}, M_{\mathcal{L}} \rightarrow \infty}\epsilon^{(\star, \mathcal{NN})} = \epsilon^{\star}. \label{eq:ref_equal_risk_price_measure_convergence}
		\end{align}
	\end{proposition}
	
	
	\subsection{Optimization of feedforward neural networks}
	\label{subsec:optim_FFNN}
	The training procedure of the long-$\mathcal{NN}$ and short-$\mathcal{NN}$ consists in searching for their optimal parameters $\theta^{(\mathcal{L})} \in \Theta_{M_{\mathcal{L}}}$ and $\theta^{(\mathcal{S})} \in \Theta_{M_\mathcal{S}}$ to minimize the measured risk exposures as in \eqref{eq:ref_risk_long_FFNN} and \eqref{eq:ref_risk_short_FFNN}. The approach utilized in this paper is based on the deep hedging algorithm of \cite{buehler2019deep}.
	The training procedure of the short-$\mathcal{NN}$ with (minibatch) stochastic gradient descent (SGD), a very popular algorithm in deep learning, is presented. It is straightforward to adapt the latter to the long-$\mathcal{NN}$ with a simple modification to the cost function \eqref{eq:ref_loss_func_def} that follows. 
	Let $J(\theta)$ be the cost function to be minimized for the short derivative position hedge\footnote{%
	Recall from \eqref{eq:ref_risk_short_FFNN} that the relation between $J(\theta)$ and the measured risk exposure of the short position is
	$$\epsilon^{(M_{\mathcal{S}})}(0) = \underset{\theta \in \Theta_{M_{\mathcal{S}}}}{\min} \, J(\theta).$$
	}:
	\begin{align}
	J(\theta):= \rho \left(\Phi(S_{N},Z_{N}) - B_{N}G_{N}^{\delta^{(S, \theta)}}\right), \quad \theta \in \Theta_{M_\mathcal{S}}. \label{eq:ref_loss_func_def}
	\end{align}
	Denote $\theta_{0} \in \Theta_{M_\mathcal{S}}$ as  the initial\footnote{
		In this paper, the initialization of $\theta$ is always done with the Glorot uniform initialization from \cite{glorot2010understanding}.
	} parameter values of $F_{\theta}^{S}$. The classical SGD algorithm consists in updating iteratively the trainable parameters as follows:
	\begin{align}
	\theta_{j+1} &= \theta_{j} - \eta_j \nabla_{\theta}J(\theta_{j}), \label{eq:ref_SGD_iterative_param}
	\end{align}
	where $\nabla_{\theta}$ denotes the gradient operator with respect to $\theta$ and $\eta_j$ is a small positive deterministic value 
	which is typically progressively reduced through iterations, i.e. $\!\!$ as $j$ increases.
	Recall that in the current framework, a synthetic market is considered where paths of the hedging instruments can be simulated. 
	Let $N_{\text{batch}} \in \mathbb{N}$ be the size of a simulated minibatch $\mathbb{B}_{j} := \{\pi_{i,j}\}_{i=1}^{N_{\text{batch}}}$ with $\pi_{i,j}$ being the $i^{\text{th}}$ hedging error if the trainable parameters are $\theta = \theta_{j}$:
	\begin{align}
	\pi_{i,j}:= \Phi(S_{N,i}, Z_{N,i}) - B_{N}G_{N,i}^{\delta^{(S,\theta_j)}}. \label{eq:ref_simulated_hedging_error}
	\end{align}  
	Moreover, let $\hat{\rho}(\mathbb{B}_{j})$ be the empirical estimator of $\rho(\pi_{i,j})$. The gradient of the cost function $\nabla_{\theta}J(\theta_j)$ is estimated with $\nabla_{\theta}\hat{\rho}(\mathbb{B}_{j})$ evaluated at $\theta = \theta_{j}$.
	
	In the numerical section, the convex risk measure is assumed to be the Conditional Value-at-Risk (CVaR) as defined in \cite{rockafellar2002conditional}. For an absolutely continuous integrable random variable\footnote{
		In \cref{sec:numerical_results}, the only dynamics considered for the risky assets produce integrable and absolutely continuous hedging errors. 
	}, the CVaR has the following representation:
	\begin{align}
	\text{CVaR}_{\alpha}(X):=\E[X|X \geq \text{VaR}_{\alpha}(X)], \quad \alpha \in (0,1), \label{eq:ref_CVaR_def}
	\end{align}
	where $\text{VaR}_{\alpha}(X) := \underset{x}{\min} \left\{x|\mathbb{P}(X \leq x) \geq \alpha \right\}$ is the Value-at-Risk (VaR) of confidence level $\alpha$. The CVaR has been extensively used in the risk management literature as it considers tail risk by averaging all losses larger than the VaR. For a simulated minibatch of hedging errors $\mathbb{B}_{j}$, let $\{\pi_{[i],j}\}_{i=1}^{N_{\text{batch}}}$ be the corresponding ordered sequence and $\tilde{N}:= \ceil*{\alpha N_{\text{batch}}}$ where $\ceil{x}$ is the ceiling function (i.e. the smallest integer greater or equal to $x$). Following the work of \cite{hong2014monte} (see Section $2$ of their paper), let $\reallywidehat{\text{VaR}}_{\alpha}(\mathbb{B}_{j})$ and $\reallywidehat{\text{CVaR}}_{\alpha}(\mathbb{B}_{j})$ be the estimators of the VaR and CVaR of the short hedging error at confidence level $\alpha$:
	$$\reallywidehat{\text{VaR}}_{\alpha}(\mathbb{B}_{j}) := \pi_{[\tilde{N}], j},$$
	$$\reallywidehat{\text{CVaR}}_{\alpha}(\mathbb{B}_{j}):= \reallywidehat{\text{VaR}}_{\alpha}(\mathbb{B}_{j}) + \frac{1}{(1-\alpha)N_{\text{batch}}}\sum_{i=1}^{N_{\text{batch}}}\max(\pi_{i,j}-\reallywidehat{\text{VaR}}_{\alpha}(\mathbb{B}_{j}),0).$$
	%
	Note that $\reallywidehat{\text{CVaR}}_{\alpha}(\mathbb{B}_{j})$ depends of every trainable parameters in $\theta_{j}$ as the $\pi_{i,j}$ are functions of the trading strategy produced by the output of the short-$\mathcal{NN}$. Furthermore, since the gain process and the trading strategy are linearly dependent, $\reallywidehat{\text{CVaR}}_{\alpha}(\mathbb{B}_{j})$ is also linearly dependent of the trading strategy. The latter implies that $\nabla_{\theta}\reallywidehat{\text{CVaR}}_{\alpha}(\mathbb{B}_{j})$ is known analytically as the gradient of the output of a FFNN with respect to the trainable parameters is known analytically (see e.g. \cite{goodfellow2016deep}).
	
	\begin{remark}
		It can be shown that $\reallywidehat{\text{CVaR}}_{\alpha}(\mathbb{B}_{j})$ is biased in finite sample size, but is a consistent and asymptotically normal estimator of the CVaR (see e.g. Theorem $2$ of \cite{trindade2007financial}). The specific impacts of this bias on the optimization procedure presented in this paper are out-of-scope. Multiple considerations
		are typically used to determine the minibatch size. It is often treated as an additional hyperparameter (see e.g. Chapter $8.1.3$ of \cite{goodfellow2016deep} for additional details). 
		Numerical results presented in \cref{sec:numerical_results} of the current paper are robust to different minibatch sizes, i.e. no significant difference was observed under different minibatch sizes.
	\end{remark}
	
	\begin{remark}
	A very popular algorithm in deep learning to compute analytically the gradient of a cost function with respect to the parameters is backpropagation \citep{rumelhart1986learning}, often called backprop. Backprop leverages efficiently the structure of neural networks and the chain rule of calculus to obtain such gradient. In practice, deep learning libraries such as Tensorflow are often used to implement backprop. Moreover, sophisticated SGD algorithms such as Adam \citep{kingma2014adam} which dynamically adapt the $\eta_j$ in \eqref{eq:ref_SGD_iterative_param} over time have been shown to improve the training of neural networks. For all of the numerical experiments in \cref{sec:numerical_results}, Tensorflow and Adam were used.
	\end{remark}
%

	\begin{remark}
		\label{remark:SGD_convergence_local_min}
		\cref{prop_ref_arbitrage_free_with_NN} shows that the current paper's approach can approximate arbitrarily well $C_{0}^{\star}$ and $\epsilon^{\star}$. As discussed in \cite{goodfellow2016deep}, SGD is not guaranteed to reach a local minimum in a reasonable amount of time. But empirically, it is often the case that the algorithm finds a very low value of the cost function quickly enough to be useful. Hence, in practice, one cannot expect the current SGD approach to converge to $C_{0}^{\star}$ and $\epsilon^{\star}$, but should instead expect to obtain close-to-optimal parameters for $F_{\theta}^{\mathcal{L}}$ and $F_{\theta}^{\mathcal{S}}$ which result in a good enough approximation.
	\end{remark}


	\section{Numerical results}
	\label{sec:numerical_results}
	This section illustrates the implementation of the equal risk pricing framework under different market setups. Our analysis starts off in \cref{subsec:different_convex_measure} with a sensitivity analyses of equal risk prices and residual hedging risk in relation with the choice of convex risk measure. The assessment of the impact of different empirical properties of assets returns on the equal risk pricing framework is performed in \cref{subsec:model_friction}. A comparison with benchmarks consisting in risk-neutral expected prices under commonly used EMMs is also presented. \cref{subsec:contingent_claim_friction} shows that the current paper's approach is very general and is able to price exotic derivatives and assess their associated residual hedging risk. The setup for the latter numerical experiments is detailed in \cref{subsec:num_procedure}.
	
	
	\subsection{Numerical procedure}
	\label{subsec:num_procedure}
	
	A single risky asset (i.e. $D = 1$) of initial price $S_{0} = 100$ is considered. It can be assumed to be a non-dividend paying stock. The annualized continuous risk-free rate is $r = 0.02$. Daily rebalancing with $260$ business days per year is applied, i.e. $t_{i} - t_{i-1} = 1/260$ for $i = 1,\ldots,N$. The contingent claim to be priced is a vanilla European put option on the risky asset with maturity $T = 60/260$. Different levels of moneyness are considered: $K=90$ for OTM, $K=100$ for at-the-money (ATM) and $K=110$ for in-the-money (ITM). The convex risk measure used by the hedger is assumed to be the CVaR risk measure. 
	
	\subsubsection{Regime-switching model}
	\label{regime_switch_model}
	For $n=1,\ldots,N$, the daily log-returns $y_{n}:= \log(S_n/S_{n-1})$ are assumed, unless stated otherwise, to follow a Gaussian regime-switching (RS) model.
	RS models have the ability to reproduce broadly accepted stylized facts of asset returns such as heteroskedasticity, autocorrelation in absolute returns, leverage effect and fat tails, see, for instance \cite{ang2012regime}. Under RS models, log-returns depend on an unobservable discrete-time process. Let $h=\{h_{n}\}_{n=0}^{N}$ be a finite state Markov chain taking values in $\{1,\ldots,H\}$ for a positive integer $H$, where $h_{n}$ is the regime or state of the market during the period $[t_{n}, t_{n+1})$.
	Let $\{\gamma_{i,j}\}_{i=1,j=1}^{H,H}$ be the homogeneous transition probabilities of the Markov chain, where for $n = 0,\ldots,N-1$\footnote{
		The distribution of $h_{0}$ is assumed to be the stationary distribution of the Markov chain.
	}:
	\begin{align}
	\mathbb{P}(h_{n+1}=j|\mathcal{F}_{n}, h_{n},\ldots,h_{0})=\gamma_{h_{n},j}, \quad j =1,\ldots,H. \label{eq:ref_transition_proba}
	\end{align}
	Let $\Delta:=1/260$. The daily log-returns are assumed to have the following dynamics:
	\begin{align}
	y_{n+1}&= \mu_{h_{n}}\Delta + \sigma_{h_{n}}\sqrt{\Delta}\epsilon_{n+1}, \quad n = 0,\ldots,N-1,\label{eq:ref_RS_model}
	\end{align}
	where $\{\epsilon_{n}\}_{n=1}^{N}$ are independent standard normal random variables and $\{\mu_{i}, \sigma_{i}\}_{i=1}^{H}$ are the yearly model parameters with $\mu_i \in \mathbb{R}$ and $\sigma_i > 0$. Following the work of \cite{godin2019option}, define $\xi:=\{\xi_{n}\}_{n=0}^{N}$ as the predictive probability process
	 with $\xi_{n}:=[\xi_{n,1},\ldots,\xi_{n,H}]$ and $\xi_{n,j}$ as the probability that the Markov chain is in the $j^{\text{th}}$ regime during $[t_{n},t_{n+1})$ conditional on the investor's filtration, i.e.:
	\begin{align}
	\xi_{n,j}:=\mathbb{P}(h_{n}=j|\mathcal{F}_{n}), \quad j = 1,\ldots,H. \label{eq:ref_filtered_proba_P}
	\end{align}
	\cite{franccois2014optimal} show that the optimal hedging portfolio composition at time $t_{n}$ is strictly a function of $\{S_{n}, V_{n}, \xi_{n}\}$. Thus, in \cref{assumpt:unique}, the feature vector considered for both the long-$\mathcal{NN}$ and short-$\mathcal{NN}$ is $X_{n} = [S_{n}, V_{n}, \xi_{n}, T - t_{n}]$. \cite{franccois2014optimal} also provide a recursion to compute the predictive probabilities processes $\xi$. For $k=1,\ldots,H$, define the function $\phi_{k}$ as the Gaussian pdf with mean $\mu_k$ and standard deviation $\sigma_k$:
	\begin{align}
	\phi_{k}(x):=\frac{1}{\sigma_{k}\sqrt{2\pi}}\exp \left(-\frac{(x - \mu_{k})^{2}}{2\sigma_{k}^{2}}\right). \nonumber
	\end{align} 
	Setting $\xi_{0}$ as the stationary distribution of the Markov chain, the $\xi_{n,i}$ can be recursively computed for $n=1,\ldots,N$ as follows:
	\begin{align}
	\xi_{n, i} &= \frac{\sum_{j=1}^{H} \gamma_{j,i} \phi_{j}(y_{n})\xi_{n-1,j}}{\sum_{j=1}^{H} \phi_{j}(y_{n})\xi_{n-1,j}}, \quad i = 1,\ldots,H. \nonumber
	\end{align}
	In \cref{subsec:model_friction}, different dynamics for the underlying will be considered. Each model is estimated with maximum likelihood on the same time series of daily log-returns on the S\&P 500 price index for the period 1986-12-31 to 2010-04-01 (5863 observations). Resulting parameters are in \cref{appendix:model_parameters}. 
	
	\subsubsection{Neural network structure}
	\label{neural_net_structure}
	The training of the long-$\mathcal{NN}$ and short-$\mathcal{NN}$ is done as described in \cref{subsec:optim_FFNN} with $100$ epochs\footnote{
		An epoch is defined as one complete iteration of SGD over the training set. For a training set of $400,\!000$ paths and a batch size of $1,\!000$, one epoch is equivalent to $400$ iterations of SGD.
	}, a minibatch size of $1,\!000$ on a training set (in-sample) of $400,\!000$ independent simulated paths and a learning rate of $0.0005$ with the Adam algorithm. 
	Numerical results presented are obtained from a test set (out-of-sample) of $100,\!000$ independent paths. The structure of every neural network is $3$ layers with $56$ neurons per layer and the activation function is the rectified linear unit (ReLU) where $\text{ReLU} : \mathbb{R} \rightarrow [0,\infty)$ is defined as $\text{ReLU}(x):=\text{max}(0,x)$.


	\subsection{Sensitivity analyses}
	\label{subsec:different_convex_measure}
	In this section, we perform sensitivity analyses of equal risk prices and residual hedging risk with respect to the confidence level of the CVaR. Three different confidence levels are considered: $\text{CVaR}_{\alpha}$ at levels $0.90, 0.95$ and $0.99$. Optimizing risk exposure using a higher level $\alpha$ corresponds to agents with a higher risk aversion as the latter puts more relative weight on losses of larger magnitude. Thus, the choice of the confidence level is motivated by the objective of assessing the impact of the level of risk aversion of hedging agents on equal risk pricing. \cref{table:sensitivity_analysis_CVAR_090_095_099} presents the equal risk option prices and residual hedging risk exposures under the three confidence levels.
	\begin{table}[ht]
		\caption {Sensitivity analysis of equal risk prices $C_{0}^{(\star, \mathcal{NN})}$ and residual hedging risk $\epsilon^{(\star, \mathcal{NN})}$ for OTM ($K=90$), ATM ($K=100$) and ITM ($K=110$) put options of maturity $T=60/260$.} \label{table:sensitivity_analysis_CVAR_090_095_099}
		\renewcommand{\arraystretch}{1.15}
		\begin{adjustwidth}{-1in}{-1in} 
			\centering
			\begin{tabular}{lccccccccccc}
				\hline\noalign{\smallskip}
				& \multicolumn{3}{c}{$C_{0}^{(\star, \mathcal{NN})}$} & & \multicolumn{3}{c}{$\epsilon^{(\star, \mathcal{NN})}$} & & \multicolumn{3}{c}{$\epsilon^{(\star,\mathcal{NN})}/C_{0}^{(\star, \mathcal{NN})}$} \\
				\cline{2-4}\cline{6-8}\cline{10-12}   Moneyness   & $\text{OTM}$ & $\text{ATM}$ & $\text{ITM}$ &   & $\text{OTM}$ & $\text{ATM}$ & $\text{ITM}$ &   & $\text{OTM}$ & $\text{ATM}$ & $\text{ITM}$ \\
				\hline\noalign{\medskip} 
				%
				$\text{CVaR}_{0.90}$ & $1.40$ & $4.19$ & $11.14$ &   & $1.36$ & $2.61$ & $1.68$ &   & $0.97$ & $0.62$ & $0.15$ \\
				$\text{CVaR}_{0.95}$ & $32\%$ & $4\%$ & $2\%$ &   & $35\%$ & $11\%$ & $14\%$ &   & $2\%$ & $7\%$ & $12\%$ \\
				$\text{CVaR}_{0.99}$ & $91\%$ & $42\%$ & $14\%$ &   & $99\%$ & $79\%$ & $98\%$ &   & $4\%$ & $26\%$ & $74\%$ \\
				\noalign{\medskip}\hline
			\end{tabular}%
		\end{adjustwidth}
		Notes: These results are computed based on $100,\!000$ independent paths generated from the regime-switching model under $\mathbb{P}$ (see \cref{regime_switch_model} for model definition and \cref{appendix:model_parameters} for model parameters). The training of neural networks is done as described in \cref{neural_net_structure}.  Values for the $\text{CVaR}_{0.95}$ and $\text{CVaR}_{0.99}$ risk measures are expressed relative to $\text{CVaR}_{0.90}$ (\% increase).
	\end{table}
	Our numerical results show that under the equal risk pricing framework, an increase in the risk aversion of hedging agents leads to increased put option prices. Indeed, under the use of the $\text{CVaR}_{0.99}$ risk measure, option prices significantly increase across all moneynesses with relative increases of respectively $91\%$, $42\%$ and $14\%$ for OTM, ATM and ITM contracts with respect to prices obtained with the $\text{CVaR}_{0.90}$. By using the $\text{CVaR}_{0.95}$ instead of $\text{CVaR}_{0.90}$, only OTM equal risk prices are significantly impacted with an increase of $32\%$, while for ATM and ITM, the increase seems marginal. The positive association between put option prices and the confidence level of hedgers can be explained by the fact that a put option payoff is bounded below by zero. Therefore, the hedging error of the short position has a much heavier right tail than for the long position. Thus, an increase in $\alpha$ often implies a larger increase for the risk exposure of the short position than for the long position. This results in a higher equal risk price to compensate the heavier increase in risk exposure of the short position.
	
	As expected, the risk exposure per option contract ($\epsilon^{(\star, \mathcal{NN})}$) also increases with the level of risk aversion across all moneynesses. This is a direct consequence of \cref{proposition:interesting_form_risk_exposure} and the monotonicity property of $\text{CVaR}_{\alpha}$ with respect to $\alpha$. Also, the risk exposure per dollar invested ($\epsilon^{(\star,\mathcal{NN})}/C_{0}^{(\star, \mathcal{NN})}$) for ITM and ATM contracts exhibits high sensitivity to the confidence level $\alpha$, while for OTM the value of $\alpha$ seems much less important. This observation for OTM contracts is due to a similar relative increase in prices and residual risk exposures obtained when $\alpha$ is increased. From these results, we can conclude that in practice, the choice of the confidence level (or more generally of the risk measure itself) needs to be carefully analyzed as it can have a material impact on equal risk option prices.
	
	
	\subsection{Model induced incompleteness}
	\label{subsec:model_friction}
	In this section, we consider four different dynamics for the underlying: the BSM, a GARCH process, a regime-switching process and a jump-diffusion. 
	This is motivated by the objective of assessing the impact of different empirical properties of asset returns on the equal risk pricing framework. Indeed, Monte Carlo simulations from these models enable quantifying the impact of time-varying volatility, regime risk and jump risk on equal risk prices and residual hedging risk.  
	Moreover, risk-neutral expected prices are used as benchmarks to equal risk prices under common EMMs found in the literature. The physical dynamics of each model is described below and the associated risk-neutral dynamics are provided in \cref{appendix:risk_neutral_dynamic}. 


	\subsubsection{Discrete BSM}
	Under the discrete Black-Scholes model, log-returns are assumed to be i.i.d. $\!\!$ normal random variables with daily mean and variance of respectively $(\mu-\frac{\sigma^{2}}{2})\Delta$ and $\sigma^{2} \Delta$:   
	\begin{align}
	y_{n}&=\left(\mu-\frac{\sigma^{2}}{2} \right)\Delta + \sigma \sqrt{\Delta}\epsilon_{n}, \quad n = 1,\ldots, N, \label{eq:ref_BSM_under_P}
	\end{align}
	where $\mu \in \mathbb{R}$ and $\sigma >0$ are the yearly model parameters, and $\{\epsilon_{n}\}^N_{n=1}$ are independent standard normal random variables. The feature vector of the neural network is $X_{n} = [S_{n}, V_{n}, T - t_{n}]$.
	
	
	\subsubsection{Discrete Merton jump-diffusion (MJD) model}
	The jump-diffusion model of \cite{Merton1976} generalizes the BSM by incorporating random jumps within paths. Let $\{\epsilon_{n}\}^N_{n=1}$ be independent standard normal random variables, $\{N_{n}\}_{n=0}^{N}$ be values of a homogeneous Poisson process of intensity $\lambda$ at $t_0,\ldots,t_N$ and $\{\rchi_{j}\}^\infty_{j=1}$ be i.i.d. $\!\!$ normal random variables of mean $\gamma \in \mathbb{R}$ and variance $\vartheta^{2}$. $\{N_{n}\}_{n=1}^{N}$, $\{\epsilon_{n}\}_{n=1}^{N}$ and $\{\rchi_{j}\}^\infty_{j=1}$ are assumed independent. For $n= 1, \ldots, N$:
	\begin{align}
	y_{n} = \left(\alpha - \lambda \left(e^{\gamma + \vartheta^{2}/2}-1\right) -  \frac{\sigma^{2}}{2}\right)\Delta + \sigma \sqrt{\Delta}\epsilon_{n} + \sum_{j=N_{n-1} + 1}^{N_{n}}\rchi_{j}, \label{eq:ref_MJD_under_P}
	\end{align}
	where $\{\alpha, \gamma, \vartheta, \lambda, \sigma\}$ are the model parameters with $\{\alpha, \lambda, \sigma\}$ being on a yearly scale, $\alpha \in \mathbb{R}$ and $\sigma > 0$. The feature vector of the neural network is $X_{n} = [S_{n}, V_{n}, T - t_{n}]$.
	

	\subsubsection{GARCH model}
	\label{subsec:GJR_GARCH}
	In constrast to the BSM or MJD model, GARCH models allow for the volatility of asset returns to be time-varying. The GJR-GARCH(1,1) model of \cite{glosten1993relation} assumes that the conditional variance of log-returns is stochastic 
%
	and captures important features of asset returns such as the leverage effect and volatility clustering. For $n=1,\ldots,N$:
	\begin{align}
	y_{n} &= \mu + \sigma_{n}\epsilon_{n}, \label{eq:ref_GARCH_P_log_ret}
	\\ \sigma_{n+1}^{2} &= \omega + \alpha \sigma_{n}^{2}(|\epsilon_{n}| - \gamma \epsilon_{n})^{2} + \beta \sigma_{n}^{2}, \nonumber
	\end{align}
	where the model parameters $\{\omega, \alpha, \beta\}$ are positive real values, $\mu \in \mathbb{R}, \gamma \in \mathbb{R}$ and $\{\epsilon_{n}\}^N_{n=1}$ is a sequence of independent standard normal random variables. 
	Given the initial value $\sigma_{1}^{2}$, $\{\sigma_{n}^{2}\}_{n=1}^{N}$ is predictable with respect to $\mathbb{F}$. 
	A common assumption which is used in this paper is to set $\sigma_{1}^{2}$ as the stationary variance: $\sigma_{1}^{2} = \frac{\omega}{1 - \alpha(1+\gamma^{2})-\beta}$. The feature vector of the neural network is $X_{n} = [S_{n}, V_{n}, \sigma_{n+1}, T - t_{n}]$. 
	

	\subsubsection{Results}
	\label{subsubsec:model_induced_incompleteness}
	\cref{table:model_induced_market_incompleteness_sec_4.2_in_percentage} presents the equal risk prices and residual hedging risk exposures for the four dynamics considered based on the $\text{CVaR}_{0.95}$ risk measure. 
	\begin{table}[ht]
	\caption {Equal risk prices $C_{0}^{(\star, \mathcal{NN})}$ and residual hedging risk $\epsilon^{(\star, \mathcal{NN})}$ for OTM ($K=90$), ATM ($K=100$) and ITM ($K=110$) put options of maturity $T=60/260$ under different dynamics.} \label{table:model_induced_market_incompleteness_sec_4.2_in_percentage}
	\renewcommand{\arraystretch}{1.15}
	\begin{adjustwidth}{-1in}{-1in} 
		\centering
		\begin{tabular}{lccccccccccc}
			\hline\noalign{\smallskip}
			& \multicolumn{3}{c}{$C_{0}^{(\star, \mathcal{NN})}$} & & \multicolumn{3}{c}{$\epsilon^{(\star, \mathcal{NN})}$} & & \multicolumn{3}{c}{$\epsilon^{(\star,\mathcal{NN})}/C_{0}^{(\star, \mathcal{NN})}$} \\
			\cline{2-4}\cline{6-8}\cline{10-12}   Moneyness   & $\text{OTM}$ & $\text{ATM}$ & $\text{ITM}$ &   & $\text{OTM}$ & $\text{ATM}$ & $\text{ITM}$ &   & $\text{OTM}$ & $\text{ATM}$ & $\text{ITM}$ \\
			\hline\noalign{\medskip} 
			%
			BSM & $0.58$ & $3.53$ & $10.39$ &   & $0.35$ & $0.74$ & $0.59$ &   & $0.60$ & $0.21$ & $0.06$ \\
			MJD & $5\%$ & $-2\%$ & $0\%$ &   & $50\%$ & $62\%$ & $41\%$ &   & $42\%$ & $65\%$ & $41\%$ \\
			GJR-GARCH & $68\%$ & $-4\%$ & $-1\%$ &   & $165\%$ & $115\%$ & $28\%$ &   & $57\%$ & $124\%$ & $30\%$ \\
			Regime-switching & $217\%$ & $23\%$ & $9\%$ &   & $428\%$ & $291\%$ & $223\%$ &   & $67\%$ & $218\%$ & $196\%$ \\
			\noalign{\medskip}\hline
		\end{tabular}%
	\end{adjustwidth}
	Notes: These results are computed based on $100,\!000$ independent paths generated from each of the four different models for the underlying under $\mathbb{P}$ (see \cref{regime_switch_model} and \cref{subsec:model_friction} for model definitions and \cref{appendix:model_parameters} for model parameters). For each model, the training of neural networks is done as described in \cref{neural_net_structure}. The confidence level of the CVaR risk measure is $\alpha = 0.95$. Results are expressed relative to the BSM (\% increase).
	\end{table}
	Values observed for $C_{0}^{(\star, \mathcal{NN})}$ indicate that the sensivity of equal risk prices with respect to the dynamics of the underlying highly depends on the moneyness. Indeed, OTM prices are significantly impacted by the choice of dynamics; choosing the GJR-GARCH or RS models instead of the BSM leads to a price increase respectively of $68\%$ or $217\%$. For ATM and ITM contracts, only the RS model seems to materially alter equal risk prices in comparison to BSM prices with respective increases of $23\%$ and $9\%$. Moreover, the numerical results confirm that the increase in hedging residual risk 
	generated by time-varying volatility, regime risk and jump risk is far from being marginal and is highly sensitive to the moneyness of the option. Values obtained for the metric $\epsilon^{(\star,\mathcal{NN})}/C_{0}^{(\star, \mathcal{NN})}$ show that regime risk has the most impact with an increase of the risk exposure per dollar invested of $67\%, 218\%$ and $196\%$ respectively for the OTM, ATM and ITM contracts in comparison to the BSM. When compared to jump risk, time-varying volatility seems to have a higher impact on residual hedging risk for OTM and ATM options, while jump risk has a higher impact on ITM contracts. It is interesting to note that \cite{augustyniak2017assessing} evaluate the impact of the dynamics of the underlying on the risk exposure and on the price of contingent claims under a quadratic penalty. Their numerical results show that the risk exposure is highly sensitive to the dynamics,
	but not the price. This is in contrast to numerical results of the current study which show that under a non-quadratic penalty, prices can also vary significantly with the dynamics of the underlying. 

%

	\begin{table}[ht]
	\caption {Equal risk and risk-neutral prices for OTM ($K=90$), ATM ($K=100$) and ITM ($K=110$) put options of maturity $T=60/260$.} \label{table:equal_risk_prices_vs_EMM}
	\renewcommand{\arraystretch}{1.15}
	\begin{adjustwidth}{-1in}{-1in} 
		\centering
		\begin{tabular}{lcccccccc}
			\hline\noalign{\smallskip}
			& \multicolumn{3}{c}{$\text{Risk-neutral prices}$} & & \multicolumn{3}{c}{$\text{Equal risk prices}$} & \\
			\cline{2-4}\cline{6-8}   Moneyness   & $\text{OTM}$ & $\text{ATM}$ & $\text{ITM}$ &   & $\text{OTM}$ & $\text{ATM}$ & $\text{ITM}$ &  \\
			\hline\noalign{\medskip} 
			%
			BSM & $0.53$ & $3.51$ & $10.36$ &  & $10\%$ & $1\%$ & $0\%$ &    \\
			MJD & $0.46$ & $3.32$ & $10.24$ &   & $34\%$ & $4\%$ & $2\%$ &  \\
			GJR-GARCH & $0.57$ & $2.98$ & $9.84$  &   & $71\%$ & $14\%$ & $4\%$ &   \\
			Regime-switching  & $0.56$ & $3.10$ & $10.33$  &   & $231\%$ & $40\%$ & $10\%$ &   \\
			\noalign{\medskip}\hline
		\end{tabular}%
	\end{adjustwidth}
	Notes: Results for equal risk prices are computed based on $100,\!000$ independent paths generated from each of the four different models for the underlying under $\mathbb{P}$ (see \cref{regime_switch_model} and \cref{subsec:model_friction} for model definitions and \cref{appendix:model_parameters} for model parameters). For each model, the training of neural networks is done as described in \cref{neural_net_structure}. The confidence level of the CVaR risk measure is $\alpha = 0.95$. Results for risk-neutral prices are computed under the associated risk-neutral dynamics described in \cref{appendix:risk_neutral_dynamic}. Equal risk prices are expressed relative to risk-neutral prices (\% increase). 
	\end{table}


	\cref{table:equal_risk_prices_vs_EMM} compares equal risk prices to risk-neutral prices for each dynamics. These results show that except for a few cases, equal risk prices are significantly higher than risk-neutral prices across all dynamics and moneynesses. This is especially true for OTM contracts: the lowest and highest relative price increases are $10\%$ and $231\%$ when going from the BSM to the regime-switching model. This significant increase in option prices can be attributed to the different treatment of market scenarios by each approach. Expected risk-neutral prices consider averages of all scenarios, while equal risk prices with the CVaR risk measure coupled with a high confidence level $\alpha$ only consider extreme scenarios. 

	The latter observation has important implications for financial participants in the option market. Indeed, by using the risk-neutral valuation approach instead of the equal risk pricing framework, a risk averse participant acting as a provider of options, e.g. a market maker, might
	significantly underprice OTM put options. From the perspective of the equal risk pricing framework, risk-neutral prices imply more residual risk for the short position of OTM put contracts than for the long position. It is important to note that the risk-neutral dynamics considered in this paper assume that jump and regime risk are not priced in the market. Additional analyses comparing equal risk prices to risk-neutral prices under alternative EMMs embedding other forms of risk premia (see for instance \cite{bates1996jumps} for jump risk premium and \cite{godin2019option} for regime risk premium) may prove worthwhile in further work.

	\subsection{Exotic contingent claims}
	\label{subsec:contingent_claim_friction}
	In this section, two exotic contingent claims are considered for the equal risk pricing framework, namely an Asian average price put and lookback put with fixed strike. For $Z_{n} = \frac{1}{n+1}\sum_{i=0}^{n} S_{i}$, $n =0,\ldots,N$, the Asian option's payoff is:
	 $$\Phi(S_{N},Z_{N}) = \max(0, K - Z_{N}).$$
	 For $Z_{n}  = \underset{i = 0,\ldots,n}{\min} S_{i}$, $n =0,\ldots,N$, the lookback option's payoff is:
	 $$\Phi(S_{N},Z_{N}) = \text{max}(0, K - Z_{N}).$$ 
	 The same assumptions as in \cref{subsec:model_friction} are imposed, and only the regime-switching model is considered. The maturity is still $T = 60/260$. The feature vector for both exotic contingent claims is $X_{n} = [S_{n}, Z_{n}, V_{n}, \xi_{n}, T-t_{n}]$. \cref{table:different_contingent_claim_sec_4.4} presents the prices and residual hedging risk for the three contingent claims (including the vanilla put option studied in previous sections) and \cref{table:equal_risk_prices_vs_EMM_exotic} compares the equal risk prices to risk-neutral prices. From \cref{table:different_contingent_claim_sec_4.4}, we observe that the residual hedging risk exposure ($\epsilon^{(\star, \mathcal{NN})}$) is the highest for the lookback option, followed by the vanilla put and then the Asian option. Although this was expected, our approach has the benefit of quantifying how risk varies across different option categories. \cref{table:equal_risk_prices_vs_EMM_exotic} shows that equal risk prices under the RS model are significantly higher than risk-neutral prices across all contingent claims considered. The difference is most important for OTM contracts with respective increases of $231\%$, $465\%$ and $220\%$ for the put, Asian and lookback options. The finding that equal risk prices tend to be higher than risk-neutral prices is therefore shared by multiple option categories.
	 
%
%
%
%

	\begin{table}[ht]
		\caption {Equal risk prices $C_{0}^{(\star, \mathcal{NN})}$ and residual hedging risk $\epsilon^{(\star, \mathcal{NN})}$ for OTM ($K=90$), ATM ($K=100$) and ITM ($K=110$) vanilla put, Asian average price put and lookback put options of maturity $T=60/260$.} \label{table:different_contingent_claim_sec_4.4}
		\renewcommand{\arraystretch}{1.15}
		\begin{adjustwidth}{-1in}{-1in} 
			\centering
			\begin{tabular}{lccccccccccc}
				\hline\noalign{\smallskip}
				& \multicolumn{3}{c}{$C_{0}^{(\star, \mathcal{NN})}$} & & \multicolumn{3}{c}{$\epsilon^{(\star, \mathcal{NN})}$} & & \multicolumn{3}{c}{$\epsilon^{(\star,\mathcal{NN})}/C_{0}^{(\star, \mathcal{NN})}$} \\
				\cline{2-4}\cline{6-8}\cline{10-12}   Moneyness   & $\text{OTM}$ & $\text{ATM}$ & $\text{ITM}$ &   & $\text{OTM}$ & $\text{ATM}$ & $\text{ITM}$ &   & $\text{OTM}$ & $\text{ATM}$ & $\text{ITM}$ \\
				\hline\noalign{\medskip} 
				%
				Put & $1.84$ & $4.34$ & $11.33$ &   & $1.84$ & $2.90$ & $1.91$ &   & $1.00$ & $0.67$ & $0.17$ \\
				Asian & $-66\%$ & $-36\%$ & $-7\%$ &   & $-65\%$ & $-32\%$ & $-55\%$ &   & $1\%$ & $6\%$ & $-52\%$ \\
				Lookback & $64\%$ & $80\%$ & $64\%$ &   & $64\%$ & $70\%$ & $205\%$ &   & $0\%$ & $-5\%$ & $86\%$ \\
				\noalign{\medskip}\hline
			\end{tabular}%
		\end{adjustwidth}
		Notes: These results are computed based on $100,\!000$ independent paths generated from the regime-switching model under $\mathbb{P}$ (see \cref{regime_switch_model} for model definition and \cref{appendix:model_parameters} for model parameters). The training of neural networks is done as described in \cref{neural_net_structure}. The confidence level of the CVaR risk measure is $\alpha = 0.95$. Results are expressed relative to the put option (\% increase).
	\end{table}

	\begin{table}[ht]
	\caption {Equal risk and risk-neutral prices for OTM ($K=90$), ATM ($K=100$) and ITM ($K=110$) vanilla put, Asian average price put and lookback put options of maturity $T=60/260$.}
	\label{table:equal_risk_prices_vs_EMM_exotic}
	\renewcommand{\arraystretch}{1.15}
	\begin{adjustwidth}{-1in}{-1in} 
		\centering
		\begin{tabular}{lcccccccc}
			\hline\noalign{\smallskip}
			& \multicolumn{3}{c}{$\text{Risk-neutral prices}$} & & \multicolumn{3}{c}{$\text{Equal risk prices}$} & \\
			\cline{2-4}\cline{6-8}   Moneyness   & $\text{OTM}$ & $\text{ATM}$ & $\text{ITM}$ &   & $\text{OTM}$ & $\text{ATM}$ & $\text{ITM}$ &  \\
			\hline\noalign{\medskip} 
			%
			Put & $0.56$ & $3.10$ & $10.33$ &   & $231\%$ & $40\%$ & $10\%$ &    \\
			Asian & $0.11$ & $1.77$ & $9.91$ &   & $465\%$ & $56\%$ & $6\%$ &  \\
			Lookback & $0.94$ & $5.61$ & $15.57$ &   & $220\%$ & $39\%$ & $19\%$ &   \\
			\noalign{\medskip}\hline
		\end{tabular}%
	\end{adjustwidth}
	Notes: Results for equal risk prices are computed with $100,\!000$ independent paths generated from the regime-switching model under $\mathbb{P}$ (see \cref{regime_switch_model} for model definition and \cref{appendix:model_parameters} for model parameters). The training of neural networks is done as described in \cref{neural_net_structure}. The confidence level of the CVaR risk measure is $\alpha = 0.95$. Results for risk-neutral prices are computed under the associated risk-neutral dynamics described in \cref{appendix:risk_neutral_dynamic}. Equal risk prices are expressed relative to risk-neutral prices (\% increase).
	\end{table}

	\section{Conclusion}
	\label{sec:conclusion}
	
	This paper presents a deep reinforcement learning approach to price and hedge financial derivatives under the equal risk pricing framework. This framework introduced by \cite{guo2017equal} sets option prices such that the optimally hedged residual risk exposure of the long and short positions in the contingent claim is equal. Adaptations to the latter scheme are used as proposed in \cite{marzban2020equal} by considering convex risk measures under the physical measure to evaluate residual risk exposures. A rigorous proof that equal risk prices under these modifications are arbitrage-free in general market settings which can include an arbitrary number of hedging instruments is given in the current paper.
	

	Moreover, a universal and tractable solution based on the deep hedging algorithm of \cite{buehler2019deep} to implement the equal risk pricing framework under very general conditions is described. Results presented in this paper, which rely on \cite{buehler2019deep}, demonstrate that our methodological approach to equal risk pricing can approximate arbitrarily well the true equal risk price. This study also introduces asymmetric $\epsilon$-completeness measures to quantify the level of unhedgeable risk associated with a position in a contingent claim. These measures complement the work of \cite{bertsimas2001hedging} who proposed market incompleteness measures under the quadratic penalty, while ours has the advantage of characterizing the risk aversion of the hedger with any convex risk measure. Additionally, the measures introduced in this paper are asymmetric in the sense that the risk for the long and short positions in the derivative are quantified by two different hedging strategies, unlike in \cite{bertsimas2001hedging} where the single variance-optimal hedging strategy is considered.
	
	Furthermore, Monte Carlo simulations were performed to study the equal risk pricing framework under a large variety of market dynamics. The behavior of equal risk pricing is analyzed through the choice of the underlying asset model and of the confidence level associated with the risk measure, and is benchmarked against expected risk-neutral pricing. The conduction of these numerical experiments crucially relied on the deep RL algorithm presented in this study. Numerical results showed that except for a few cases, equal risk prices are significantly higher than risk-neutral prices across all dynamics and moneynesses considered. This finding is shown to be most important for OTM contracts and shared by multiple option categories. Furthermore, for a fixed model for the underlying, sensitivity analyzes show that the choice of confidence level under the CVaR risk measure has a material impact on equal risk prices. Numerical experiments also provided insight on drivers of the $\epsilon$-completeness measures introduced in the current paper. The numerical study confirms that for vanilla put options, the increase in hedging residual risk generated by time-varying volatility, regime risk and jump risk is far from being marginal and is highly sensitive to the moneyness of the option.
	
	Future research on equal risk pricing could prove worthwhile. First, a question which remains is whether the consistence of equal risk pricing approach with risk-neutral valuations can be made explicit. Moreover, additional analyses comparing equal risk prices to risk-neutral prices under alternative EMMs embedding other forms of risk premia may also prove worthwhile. Furthermore, a numerical study of the equal risk pricing framework under other convex measures than the CVaR could be of interest. We note that \cite{marzban2020equal} provide numerical results of equal risk pricing under the worst-case risk measure in the context of robust optimization. Lastly, the financial market could be extended by including different market frictions such as transaction costs and trading constraints. The latter inclusions would require examining if equal risk prices are guaranteed to remain arbitrage-free in this context.

	\bibliographystyle{apalike}
	\bibliography{Biblio}

	
	\appendix
	
	
	 
	\section{Proofs}\label{appendix:proofs}

	
	\subsection{Proof of \texorpdfstring{\cref{prop:risk_exposure}}{Proposition \ref{prop:risk_exposure}}}
	Using the translation invariance property of $\rho$,
	\begin{align}
	\epsilon^{(L)}(V_0)&= \underset{\delta\in \Pi}{\min} \, \rho \left(- \Phi(S_{N},Z_{N})-B_{N}(V_0 + G_{N}^{\delta})\right) = \underset{\delta\in \Pi}{\min} \, \rho \left(- \Phi(S_{N},Z_{N})-B_{N}G_{N}^{\delta}\right) - B_{N}V_{0}  \nonumber 
	\\ & = \epsilon^{(L)}(0) - B_{N}V_{0}. \nonumber
	\end{align}
	Similar steps show that $\epsilon^{(S)}(V_0) = \epsilon^{(S)}(0) - B_{N}V_{0}. \quad \square$

	
	\subsection{Proof of \texorpdfstring{\cref{prop:ind_V0_hedge}}{Proposition \ref{prop:ind_V0_hedge}}}
	Using the translation invariance property of $\rho$,
	\begin{align}
	\delta^{(L)}&= \underset{\delta \in \Pi}{\argmin} \, \rho \left(- \Phi(S_{N},Z_{N})-B_{N}(V_{0} + G_{N}^{\delta})\right) =  \underset{\delta \in \Pi}{\argmin} \, \left\{\rho \left(- \Phi(S_{N},Z_{N})-B_{N}G_{N}^{\delta}\right) - B_{N}V_{0}\right\}\nonumber
	\\ &= \underset{\delta \in \Pi}{\argmin} \, \rho \left(- \Phi(S_{N},Z_{N})-B_{N}G_{N}^{\delta}\right). \nonumber
	\end{align}
	Similar steps show that $\delta^{(S)}$ is independent of $V_{0}$. $\quad \square$
	
	
	\subsection{Lemma 1} \label{subsec:lemma_1}
	\label{le:closed_trading_set}
	\textit{For any $\bar{\delta}, \tilde{\delta} \in \Pi$, $G_{n}^{\bar{\delta}+\tilde{\delta}} = G_{n}^{\bar{\delta}}+G_{n}^{\tilde{\delta}}$ for $n=0,\ldots,N$.}

	
	\subsection{Proof of Lemma 1}
	$G_{0}^{\bar{\delta}+\tilde{\delta}} = G_{0}^{\bar{\delta}}+G_{0}^{\tilde{\delta}}=0$ by definition and for $n=1,\ldots, N$:
	$$G_{n}^{\bar{\delta}+\tilde{\delta}} = \sum_{k=1}^{n} (\bar{\delta}_{k}^{(1:D)} + \tilde{\delta}_{k}^{(1:D)}) \bigcdot (B_{k}^{-1} S_{k} - B_{k-1}^{-1} S_{k-1}) =  G_{n}^{\bar{\delta}} + G_{n}^{\tilde{\delta}}. \quad \square$$


	\subsection{Lemma 2} \label{subsec:lemma_2}
	\label{le:gain_long_short}
	\textit{For any $\delta \in \Pi$, $G_{n}^{\delta} = -G_{n}^{-\delta}$ for $n = 0,\ldots,N$.}
	


	\subsection{Proof of Lemma 2}
	For $n=0$, the result is trivial as $G_{0}^{\delta}=0$ for any $\delta \in \Pi$. For $n=1,\ldots, N$:
	$$	G_{n}^{\delta}= \sum_{k=1}^{n} \delta_{k}^{(1:D)} \bigcdot (B_{k}^{-1} S_{k} - B_{k-1}^{-1} S_{k-1}) = -\sum_{k=1}^{n} (-\delta_{k}^{(1:D)}) \bigcdot (B_{k}^{-1} S_{k} - B_{k-1}^{-1} S_{k-1}) = -G_{n}^{-\delta}. \quad \square$$

	
	\subsection{Proof of \texorpdfstring{\cref{theorem:uniqueness_equal_risk_price}}{Proposition \ref{theorem:uniqueness_equal_risk_price}}}
	Using the results of \cref{prop:risk_exposure}: 
	\begin{align}
	\epsilon^{(L)}(-C_{0}^{\star}) &= \epsilon^{(S)}(C_{0}^{\star}) \Longleftrightarrow \epsilon^{(L)}(0) + B_{N}C_{0}^{\star} = \epsilon^{(S)}(0) - B_{N}C_{0}^{\star} \Longleftrightarrow C_{0}^{\star} = \frac{\epsilon^{(S)}(0) - \epsilon^{(L)}(0)}{2 B_{N}}. \label{eq:ref_form_eq_risk_price}
	\end{align}
	This shows that $C_{0}^{\star}$ exists, is unique and is given by \eqref{eq:ref_equalriskprice}. Next, we show that the equal risk price is arbitrage-free. Some parts of the proof are inspired by the work of \cite{xu2006risk}. Let $(\bar{v}, \bar{\delta})$ be a super-replication strategy of $\Phi$, see \cref{def:super_sub_strat}, where $\bar{v}$ is the super-replication price as in \eqref{eq:ref_super_replication_price}.
	Let $\tilde{\delta}:= \underset{\delta \in \Pi}{\argmin}  \, \rho\left(- B_{N}G_{N}^{\delta}\right)$.
	Using the translation invariance and monotonicity properties of $\rho$ and Lemma $1$:
	\begin{align}
	\epsilon^{(S)}(0) &= \underset{\delta \in \Pi}{\text{min }} \rho\left(\Phi(S_{N},Z_{N}) - B_{N}G_{N}^{\delta}\right) \nonumber
	\\ & \leq \rho\left(\Phi(S_{N},Z_{N}) - B_{N}(G_{N}^{\bar{\delta}+\tilde{\delta}})\right) \nonumber
	\\ &= \rho\left(\Phi(S_{N},Z_{N}) - B_{N}(G_{N}^{\bar{\delta}} + G_{N}^{\tilde{\delta}})\right) \nonumber
	\\ &= \rho\left(\Phi(S_{N},Z_{N}) - B_{N}(\bar{v} + G_{N}^{\bar{\delta}}) - B_{N}G_{N}^{\tilde{\delta}}\right) + B_{N}\bar{v} \nonumber
	\\ & \leq \rho(-B_{N}G_{N}^{\tilde{\delta}}) + B_{N}\bar{v}, \label{eq:ref_proof_short_4}
	\end{align}
	where for \eqref{eq:ref_proof_short_4}, the monotonicity property is applied to $\Phi(S_{N},Z_{N}) - B_{N}(\bar{v} + G_{N}^{\bar{\delta}}) \leq 0$ $\mathbb{P}$-a.s. This implies
	\begin{align}
	\zeta^{(S)}:= \frac{\epsilon^{(S)}(0) - \rho(-B_{N}G_{N}^{\tilde{\delta}})}{B_{N}} \leq \bar{v}. \label{eq_ref_5}
	\end{align}
	Similarly, let $(\underline{v}, \underline{\delta})$ be a sub-replication strategy where $\underline{v}$ is the sub-replication price. 
	Using the translation invariance and monotonicity properties of $\rho$ as well as Lemma $1$ and Lemma $2$:
	\begin{align}
	\epsilon^{(L)}(0)&= \underset{\delta \in \Pi}{\text{min }} \rho\left(-\Phi(S_{N},Z_{N}) -B_{N}G_{N}^{\delta}\right)  \nonumber
	\\ &\leq \rho\left(-\Phi(S_{N},Z_{N}) - B_{N}G_{N}^{\tilde{\delta} -\underline{\delta}}\right) \nonumber
	\\ &= \rho\left(B_{N}G_{N}^{\underline{\delta}} -\Phi(S_{N},Z_{N}) -  B_{N}G_{N}^{\tilde{\delta}} \right) \nonumber
	\\ &= \rho\left(B_{N}(\underline{v}+G_{N}^{\underline{\delta}}) -\Phi(S_{N},Z_{N}) - B_{N}G_{N}^{\tilde{\delta}}\right) - B_{N}\underline{v} \nonumber
	\\ &\leq \rho(-B_{N}G_{N}^{\tilde{\delta}}) - B_{N}\underline{v},  \label{eq:ref_proof_long_4}
	\end{align}
	where for \eqref{eq:ref_proof_long_4}, the monotonicity property is applied to $B_{N}(\underline{v} + G_{N}^{\underline{\delta}}) - \Phi(S_{N},Z_{N})\leq 0$ $\mathbb{P}$-a.s. This implies
	\begin{align}
	\underline{v} \leq \zeta^{(L)}:=\frac{\rho(-B_{N}G_{N}^{\tilde{\delta}})  - \epsilon^{(L)}(0)}{B_{N}}. \label{eq_ref_10}
	\end{align}

	Using \eqref{eq:ref_form_eq_risk_price}, $C_{0}^{\star}$ has the representation $C_{0}^{\star} = 0.5(\zeta^{(L)} + \zeta^{(S)})$. The last step of the proof entails showing that $\zeta^{(L)} \leq \zeta^{(S)}$, which implies that the derivative price $C_{0}^{\star} \in [\underline{v}, \bar{v}]$ and is arbitrage-free in the sense of \cref{def:arb_free_pricing}. Define the risk measure $\varrho$ as 
	\begin{align}
	\varrho(X):= \underset{\delta \in \Pi}{\text{min }} \rho\left(X - B_{N}G_{N}^{\delta}\right). \label{eq:ref_convex_risk_func}
	\end{align}
	\cite{buehler2019deep} show that since $\rho$ is a convex risk measure and $\Pi$ is a convex set, $\varrho$ is a convex risk measure (see Proposition $3.1$ of their paper).
	Note that $\epsilon^{(S)}(0) = \varrho(\Phi(S_{N},Z_{N}))$ and $\epsilon^{(L)}(0) = \varrho(-\Phi(S_{N},Z_{N}))$ by definition.
	With the translation invariance and convexity properties of $\varrho$, we obtain that
	\begin{align}
	\underset{\delta \in \Pi}{\text{min }} \rho\left(- B_{N}G_{N}^{\delta}\right) &= \varrho(0) = \varrho\left(\frac{1}{2}\Phi(S_{N},Z_{N}) - \frac{1}{2}\Phi(S_{N},Z_{N})\right)\nonumber
	\\ & \leq \frac{1}{2}\varrho(\Phi(S_{N},Z_{N})) + \frac{1}{2}\varrho(-\Phi(S_{N},Z_{N})) \nonumber
	\\ &= \frac{1}{2} \epsilon^{(S)}(0) + \frac{1}{2} \epsilon^{(L)}(0) \nonumber
	\\ \Longrightarrow \frac{\underset{\delta \in \Pi}{\text{min }} \rho\left(- B_{N}G_{N}^{\delta}\right) - \epsilon^{(L)}(0)}{B_{N}} & \leq \frac{\epsilon^{(S)}(0) - \underset{\delta \in \Pi}{\text{min }} \rho\left(- B_{N}G_{N}^{\delta}\right)}{B_{N}} \nonumber
	\\ \Longrightarrow \zeta^{(L)} &\leq \zeta^{(S)}. \quad \square \notag
	\end{align}

	
	\subsection{Proof of \texorpdfstring{\cref{proposition:interesting_form_risk_exposure}}{Proposition \ref{proposition:interesting_form_risk_exposure}}}
		Consider the equal risk price $C_{0}^{\star}$ defined in \eqref{eq:ref_what_ever}. By the definition of $\epsilon^{\star}$, \cref{prop:risk_exposure} and \cref{theorem:uniqueness_equal_risk_price}:
		$$\epsilon^{\star} = \epsilon^{(L)}(-C_{0}^{\star}) = \epsilon^{(L)}(0) + B_{N}C_{0}^{\star} = \epsilon^{(L)}(0) + B_{N}\left(\frac{\epsilon^{(S)}(0) - \epsilon^{(L)}(0)}{2B_{N}}\right) = \frac{\epsilon^{(L)}(0) + \epsilon^{(S)}(0)}{2}. \quad \square$$
	

	\subsection{Proof of \texorpdfstring{\cref{prop_ref_arbitrage_free_with_NN}}{Proposition \ref{prop_ref_arbitrage_free_with_NN}}}
		This is a direct consequence of Proposition $4.3$ in \cite{buehler2019deep} as stated in equation \eqref{eq:ref_convergence} of the current paper applied to $C_{0}^{(\star, \mathcal{NN})}$ and $\epsilon^{(\star, \mathcal{NN})}$ as stated in \eqref{eq:ref_equal_risk_price_measure_form}. $\quad \square$


	\section{Risk-neutral dynamics}
	\label{appendix:risk_neutral_dynamic}

	Since the market is arbitrage-free under the models assumed for the underlying, 
	the first fundamental theorem of asset pricing implies that their exist a probability measure $\mathbb{Q}$ such that $\{S_{n}e^{-r t_{n}}\}_{n=0}^{N}$ is an ($\mathbb{F}, \mathbb{Q}$)-martingale (see, for instance, \cite{delbaen1994general}). For the rest of \cref{appendix:risk_neutral_dynamic}, let $\{\epsilon_{n}^{\mathbb{Q}}\}_{n=1}^{N}$ be independent standard normal random variables under $\mathbb{Q}$ and denote $P_{0,T}$ as the price at time $0$ of a contingent claim of payoff $\Phi(S_{N}, Z_{N})$ at maturity $T$:
	\begin{align}
	P_{0,T} &:= e^{-rT}\E^{\mathbb{Q}}\left[\Phi(S_{N}, Z_{N}) \bigm| \mathcal{F}_{0}\right]. \label{eq:ref_put_option_price}
	\end{align} 
	Here are the risk-neutral dynamics for each model considered.
	
	
	\subsection{Regime-switching}
	\label{appendix_RS_Gaussian_risk_neutral}
	The change of measure considered is the so-called regime-switching mean-correcting transform, a popular choice under RS models (see, e.g. \cite{hardy2001regime} and \cite{bollen1998valuing}). This change of measure 
	preserves the model dynamics of regime-switching except for a shift to the drift in each respective regime. More precisely, during the passage from $\mathbb{P}$ to $\mathbb{Q}$, the transition probabilities of the Markov chain and the volatilities are left unchanged, but the drifts $\mu_{i}\Delta$ are shifted to $(r - \sigma_{i}^{2}/2)\Delta$ for regimes $i=1,\ldots,H$. 
	The resulting dynamics for the log-returns under $\mathbb{Q}$ is
	\begin{align}
	y_{n+1}&= \left(r - \frac{\sigma_{h_{n}}^{2}}{2}\right)\Delta + \sigma_{h_{n}}\sqrt{\Delta}\epsilon_{n+1}^{\mathbb{Q}}, \quad n = 0,\ldots,N-1.\label{eq:ref_RS_model_under_Q}
	\end{align}
	Let $\mathbb{H}:=\{\mathcal{H}_{n}\}_{n=0}^{N}$ 
	be the filtration generated by the markov chain $h$: 
	\begin{align} 
	\mathcal{H}_{n}:= \sigma(h_{n} \bigm| u=0,\ldots,n), \quad n = 0,\ldots,N. \label{eq:ref_filtration_regimes} 
	\end{align}
	Following the work of \cite{godin2019option}, option prices can be developed as follow. Let $\mathbb{G}:=\{\mathcal{G}_{n}\}_{n=0}^{N}$ be the filtration which contains all latent factors as well as information available to market participants, i.e. $\mathbb{G} = \mathbb{F} \vee \mathbb{H}$. Thus, the process $\{(S_{n}, h_{n})\}$ is Markov under $\mathbb{Q}$ with respect to $\mathbb{G}$. With the law of iterated expectations, the time-$0$ price of a derivative $P_{0,T}$ can be written as follows:
	\begin{align}
	P_{0,T} &=e^{-rT}\E^{\mathbb{Q}}\left[\Phi(S_{N}, Z_{N}) \bigm| \mathcal{F}_{0}\right] \nonumber
	\\ &= e^{-rT}\E^{\mathbb{Q}}\left[\E^{\mathbb{Q}}\left[\Phi(S_{N}, Z_{N}) \bigm| \mathcal{G}_{0}\right] \bigm| \mathcal{F}_{0} \right] \nonumber
	\\ &= e^{-rT}\sum_{j=1}^{H} \xi_{0,j}^{\mathbb{Q}} \E^{\mathbb{Q}}\left[\Phi(S_{N}, Z_{N}) \bigm| S_{0}, h_{0} = j\right], \label{eq:ref_RS_vasicek_price}
	\end{align}
	where $\xi_{0}^{\mathbb{Q}}$ is assumed to be the stationary distribution of the Markov chain under $\mathbb{P}$. The computation of $P_{0,T}$ can be done through Monte Carlo simulations for all contingent claims (i.e. vanilla and exotic).
	
	
	\subsection{Discrete BSM}
	\label{subsec:BSM_model_under_Q}
	By a discrete-time version of the Girsanov theorem, there exists a market price of risk process $\{\tilde{\lambda}_{n}\}_{n=1}^{N}$ such that $\epsilon_{n}^{\mathbb{Q}} = \epsilon_{n} + \tilde{\lambda}_{n}$, for $n=1,\ldots,N$. Setting $\tilde{\lambda}_{n} := \sqrt{\Delta}\left(\frac{\mu-r}{\sigma}\right)$ and replacing $\epsilon_{n} = \epsilon_{n}^{\mathbb{Q}}  - \tilde{\lambda}_{n}$ into \eqref{eq:ref_BSM_under_P}, it is straightforward to obtain the $\mathbb{Q}$-dynamics of the log-returns:
	\begin{align}
	y_{n} &= \left(r-\frac{\sigma^{2}}{2} \right)\Delta + \sigma \sqrt{\Delta}\epsilon_{n}^{\mathbb{Q}}, \quad n = 1,\ldots, N. \label{eq:ref_BSM_under_Q}
	\end{align}
	The computation of $P_{0,T}$ can be done with the well-known closed-form solution for vanilla put options (i.e. the Black-Scholes equation) and through Monte Carlo simulations for exotic contingent claims.
	
	
	\subsection{Discrete MJD}
	The change of measure used assumes no risk premia for jumps as in \cite{Merton1976} 
	and simply shifts the drift in \eqref{eq:ref_MJD_under_P} from $\alpha$ to $r$. The $\mathbb{Q}$-dynamics is thus
	\begin{align}
	y_{n} = \left(r - \lambda \left(e^{\gamma + \delta^{2}/2}-1\right) -  \frac{\sigma^{2}}{2}\right)\Delta + \sigma \sqrt{\Delta}\epsilon_{n}^{\mathbb{Q}} + \sum_{j=N_{n-1} +1}^{N_{n}}\rchi_{j}, \nonumber
	\end{align}
	where $\{\rchi_{j}\}_{j=1}^{\infty}$ and $\{N_{n}\}_{n=1}^{N}$ have the same distribution than under $\mathbb{P}$. The computation of $P_{0,T}$ for vanilla put options can be quickly performed with the fast Fourier transform (see, e.g. \cite{carr1999option}). The pricing of exotic contingent claims can be done through Monte Carlo simulations.
	
	
	\subsection{GARCH}
	\label{appendix_GJR_GARCH_risk_neutral}
	The risk-neutral measure considered is often used in the GARCH option pricing literature under which the one-period ahead conditional log-return mean is shifted, but the one-period ahead conditional variance is left untouched (see e.g. \cite{duan1995garch}). For $n=1,\ldots, N$, let $\varphi_{n} \in \mathcal{F}_{n-1}$ and define $\epsilon_{n}^{\mathbb{Q}}:= \epsilon_{n} + \varphi_{n}$. Replacing $\epsilon_{n} = \epsilon_{n}^{\mathbb{Q}} - \varphi_{n}$ into \eqref{eq:ref_GARCH_P_log_ret}, we obtain $y_{n} = \mu - \sigma_{n}\varphi_{n} + \sigma_{n}\epsilon_{n}^{\mathbb{Q}}$ for $n=1,\ldots,N$. To ensure that $\{S_{n}e^{-r t_{n}}\}_{n=0}^{N}$ is an ($\mathbb{F}, \mathbb{Q}$)-martingale, the one-period conditional expected return under $\mathbb{Q}$ must be equal to the daily risk-free rate, i.e.:
	\begin{align}
	\E^{\mathbb{Q}}[e^{y_{n}} \bigm| \mathcal{F}_{n-1}] = e^{\mu - \sigma_{n}\varphi_{n} + \sigma_{n}^{2}/2} = e^{r \Delta} \Longleftrightarrow \varphi_{n}:=\frac{\mu - r \Delta + \sigma_{n}^{2}/2}{\sigma_{n}}, \quad n = 1,\ldots,N. \nonumber
	\end{align}
	Thus, the $\mathbb{Q}$-dynamics of the GJR-GARCH(1,1) model is:
	\begin{align}
	y_{n} &= r \Delta - \sigma_{n}^{2}/2 + \sigma_{n}\epsilon_{n}^{\mathbb{Q}}, \nonumber
	\\ \sigma_{n+1}^{2} &= \omega + \alpha \sigma_{n}^{2}(|\epsilon_{n}^{\mathbb{Q}} - \varphi_{n}| - \gamma(\epsilon_{n}^{\mathbb{Q}}-\varphi_{n}))^{2} + \beta \sigma_{n}^{2}. \nonumber
	\end{align}
	The computation of $P_{0,T}$ can be done through Monte Carlo simulations for all contingent claims.

	
	\section{Maximum likelihood estimates results}
	\label{appendix:model_parameters}
	
	This section presents estimated parameters for the various underlying asset models considered in numerical experiments from \cref{sec:numerical_results}.

		\begin{table}[!htbp]
		\caption {Maximum likelihood parameter estimates of the Black-Scholes model.} \label{table:MLE_SSP500_19861231_20100401_all_models}
		\begin{adjustwidth}{-1in}{-1in} 
			\centering
			\begin{tabular}{cc}
				\hline
				$\mu$ & $\sigma$
				\\
				\hline\noalign{\medskip}
				$0.0892$  &  $0.1952$
				\\    
				\noalign{\medskip}\hline
			\end{tabular}%
		\end{adjustwidth}
		Notes: Parameters were estimated on a time series of daily log-returns on the S\&P 500 index for the period 1986-12-31 to 2010-04-01 (5863 log-returns). Both $\mu$ and $\sigma$ are on an annual basis.
		\vspace{1.1cm}
		
		\caption {Maximum likelihood parameter estimates of the GJR-GARCH(1,1) model.}
		\begin{adjustwidth}{-1in}{-1in} 
			\centering
			\begin{tabular}{ccccc}
				\hline
				$\mu$ & $\omega$ & $\alpha$ & $\gamma$ & $\beta$
				\\
				\hline\noalign{\medskip}
				%
				$2.871\text{e-}04$  &  $1.795\text{e-}06$ & $0.0540$ & $0.6028$ & $0.9105$
				\\    
				\noalign{\medskip}\hline
			\end{tabular}%
		\end{adjustwidth}
		Notes: Parameters were estimated on a time series of daily log-returns on the S\&P 500 index for the period 1986-12-31 to 2010-04-01 (5863 log-returns).
		
		\vspace{1.1cm}
		
		\caption {Maximum likelihood parameter estimates of the regime-switching model.}
		\begin{adjustwidth}{-1in}{-1in} 
			\centering
			\begin{tabular}{cccc}
				\hline
				& & $\text{Regime}$ & 
				\\			
				\hline
				$\text{Parameter}$ & $1$ & $2$ & $3$
				\\ 
				$\mu$ & $0.2040$ & $0.0337$ & $-0.6168$
				\\    
				$\sigma$ & $0.0971$ & $0.1865$ & $0.5070$
				\\
				$\nu$ & $0.4755$ & $0.4561$ & $0.0684$
				\\
				\hline
				& $0.9870$ & $0.0127$ & $0.0003$
				\\
				$\gamma$ & $0.0139$ & $0.9807$ & $0.0053$          
				\\ %
				& $0.0000$ & $0.0380$ & $0.9620$         
				\\    
				\noalign{\medskip}\hline
			\end{tabular}%
		\end{adjustwidth}
		Notes: Parameters were estimated with the EM algorithm of \cite{dempster1977maximum} on a time series of daily log-returns on the S\&P 500 index for the period 1986-12-31 to 2010-04-01 (5863 log-returns). $\nu$ represent probabilities associated with the stationary distribution of the Markov chain. $\gamma$ is the transition matrix as in \eqref{eq:ref_transition_proba}. $\mu$ and $\sigma$ are on an annual basis.
		
		\vspace{1.1cm}
		
		\caption {Maximum likelihood parameter estimates of the Merton jump-diffusion model.} 
		\begin{adjustwidth}{-1in}{-1in} 
			\centering  
			\begin{tabular}{ccccc}
				\hline
				$\alpha$ & $\sigma$ & $\lambda$ & $\gamma$ & $\vartheta$
				\\
				\hline\noalign{\medskip}
				$0.0875$  &  $0.1036$ & $92.3862$ & $-0.0015$ & $0.0160$
				\\    
				\noalign{\medskip}\hline
			\end{tabular}%
		\end{adjustwidth}
		Notes: Parameters were estimated on a time series of daily log-returns on the S\&P 500 index for the period 1986-12-31 to 2010-04-01 (5863 log-returns). $\alpha$, $\sigma$ and $\lambda$ are on an annual basis.
	\end{table}
	

	\end{document}